\documentclass[pdftex,twocolumn,3]{jour3}          

\usepackage[utf8]{inputenc}
\usepackage{epsf}
\usepackage{latexsym,amssymb,euscript}
\usepackage{widetext}
\usepackage[dvips]{graphicx}
\usepackage[numbers,sort&compress]{natbib}
\usepackage{etoolbox}
\usepackage{amsmath}
\usepackage{nicefrac}
\usepackage{slashed}
\usepackage{booktabs}
\usepackage[linktocpage]{hyperref}
\usepackage{braket}
\usepackage{chngcntr}
\usepackage{bm}
\usepackage{bbold}
\usepackage{multicol}
\usepackage{graphics}
\usepackage{graphicx}
\usepackage{mciteplus}
\usepackage{caption}
\usepackage{subcaption}
\usepackage{pdfpages}
\usepackage[titletoc]{appendix}
\usepackage[normalem]{ulem}
\usepackage{microtype} 

\graphicspath{{./figures/}}
\hypersetup{
 linktocpage = false,
 urlcolor = urlblue,
 colorlinks = true,
 linkcolor = citegreen,
 anchorcolor = linkred,
 citecolor = purple,
 pdfstartview = {XYZ null null 1.25} 
           }
\usepackage[left=2cm, right=2cm]{geometry}
\usepackage{pstricks}
\usepackage{color}
\usepackage{xcolor}
\definecolor{urlblue}{rgb}{0.2,0.4,0.7}
\definecolor{citegreen}{rgb}{0,0.4,0.2}
\definecolor{linkred}{rgb}{0.9,0.2,0.1}
\usepackage{float}
\usepackage{academicons}
\definecolor{orcidlogocol}{HTML}{A6CE39}
\usepackage{fancyhdr}
\newcommand{\drv}{{\rm d}}

\newcommand{\as}{\alpha_s}

\newcommand{\MSb}{\overline{\rm MS}}

\newcommand{\LL}{{\rm LL/LO}}

\newcommand{\NLLp}{{\rm NLL/NLO^+}}

\newcommand{\HENLOp}{{\rm HE}\mbox{-}{\rm NLO^+}}

\newcommand{\CkLL}{{\cal C}_k^\LL}

\newcommand{\CkNLLp}{{\cal C}_k^\NLLp}

\newcommand{\CkHENLOp}{{\cal C}_k^{{\rm HE}\text{-}{\rm NLO}^+}}

\newcommand{\DY}{\Delta Y}

\newcommand{\vqTTa}{\langle {\vec q}_T^{\;2} \rangle}

\newcommand{\E}{{\cal E}}

\newcommand{\HQ}{{\cal H}_Q}

\newcommand{\Jpsi}{J/\psi}

\newcommand{\BCs}{B_c(^1S_0)}
\newcommand{\Bss}{B_c(^3S_1)}

\newcommand{\XQq}{X_{Qq\bar{Q}\bar{q}}}

\newcommand{\TQQ}{T_{4Q}}

\newcommand{\TQc}{T_{4c}}

\newcommand{\TQcTpp}{T_{4c}(2^{++})}

\newcommand{\PQc}{P_{5c}}

\newcommand{{\HFNRevo}}{\tt HF-NRevo}

\newcommand{{\Jethad}}{\tt JETHAD}
\newcommand{{\symJethad}}{\tt symJETHAD}
\newcommand{{\psymJethad}}{\tt (sym)JETHAD}
\newcommand{{\Hell}}{\tt HELL}
\newcommand{{\RadISH}}{\tt RadISH}
\newcommand{{\Pegasus}}{\tt QCD-PEGASUS}
\newcommand{{\HOPPET}}{\tt HOPPET}
\newcommand{{\QCDNUM}}{\tt QCDNUM}
\newcommand{{\APFEL}}{\tt APFEL}
\newcommand{{\APFELpp}}{\tt APFEL++}
\newcommand{{\APFELppp}}{\tt APFEL(++)}
\newcommand{{\EKO}}{\tt EKO}
\newcommand{{\FeynCalc}}{\tt FeynCalc}

\newcommand{\orcidFGC}{\href{https://orcid.org/0000-0003-3299-2203}{\includegraphics[scale=0.1]{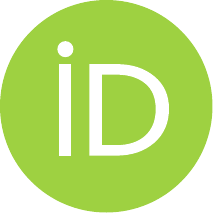}}}

\smartqed  

\journalname{}

\begin{document}

\normalem

\title{\huge Unwinding the rare $\rm \Omega$ sector: Fragmentation of \\[0.075cm] fully charmed baryons from HL-LHC to FCC
}

\subtitle{}

\author{
Francesco Giovanni Celiberto
\thanksref{e1,addr1} \orcidFGC
}

\thankstext{e1}{{\it e-mail}:
\href{mailto:fceliberto@ectstar.eu}{francesco.celiberto@uah.es}}

\institute{Universidad de Alcal\'a (UAH), E-28805 Alcal\'a de Henares, Madrid, Spain\label{addr1}
}

\date{\today}

\maketitle


\section*{Abstract}
By adopting a hadron-structure-oriented approach, we present and discuss the release of the novel {\tt OMG3Q1.0} set of collinear fragmentation functions for fully charm\-ed, rare ${\rm \Omega}$ baryons.
Our methodology combines diquarklike proxy model inputs for both charm-quark and gluon channels, calculated at the initial energy scales, with a DGLAP evolution that ensures a consistent treatment of heavy-quark thresholds, following directly from the {\HFNRevo} scheme.
We complement our work with a phenomenological study of NLL/NLO$^+$ resum\-med ${\rm \Omega}_{3c}$ plus jet distributions using {\tt (sym)JETHAD} at the HL-LHC and the future FCC.
Unraveling the production mechanisms of rare, yet-unobserved hadrons, as provided by the {\tt OMG3Q1.0} functions, stands as a key asset for deepening our understanding of QCD at future high-energy hadron colliders.
\vspace{0.30cm} \hrule
\vspace{0.30cm}
{
 \setlength{\parindent}{0pt}
 \textsc{Keywords}: \vspace{0.00cm} \\ Hadronic structure, Precision QCD, \\ Heavy flavor, Rare baryons, Omega sector, \\ Fragmentation, Resummation, \\ {\tt OMG3Q1.0} FF release
}
\vspace{-0.55cm}

\setcounter{tocdepth}{3}
\renewcommand{\baselinestretch}{1.2}\normalsize
\tableofcontents
\renewcommand{\baselinestretch}{1.0}\normalsize

\clearpage


\section{Hors d'{\oe}uvre}
\label{sec:introduction}

Heavy-flavored hadrons stand at the forefront of the search for New Physics.
As natural gateways to potential Beyond the Standard Model (BSM) interactions, they serve as probes of unknown forces, thanks to their expected coupling to hypothetical particles beyond the Standard Model spectrum. 
Their unique sensitivity to rare processes and symmetry violations renders them prime observables in high-precision experiments and theoretical studies.

At the same time, the study of heavy flavors plays a central role in our quest to decode the microscopic structure of matter. 
It offers an unparalleled window into the dynamics of strong interactions, bridging perturbative and nonperturbative regimes of Quantum Chromodynamics (QCD). 
In this context, triply heavy baryons, such as the ${\rm \Omega}_{3c}$ and ${\rm \Omega}_{3b}$, occupy a unique position in the hadronic spectrum predicted by QCD. 
Composed entirely of charm or bottom valence quarks, these particles are free from light-quark contributions, and thus provide an exceptional laboratory to probe the dynamics of color confinement in the heavy sector~\cite{Bjorken:1985ei,Fleck:1989mb,Martynenko:2007je,Martynenko:2013eoa,Karliner:2014gca,Yoshida:2015tia}. 
As color-singlet bound states of three heavy quarks, their masses, decay widths, production mechanisms, and potential observability are the subject of intense theoretical scrutiny~\cite{Ebert:2002ig,Roberts:2007ni,Chen:2011mb,Padmanath:2013zfa,Brown:2014ena,Meinel:2012qz}. 

While numerous studies have primarily addressed baryons containing two heavy quarks, triply heavy systems like the ${\rm \Omega}_{3c}$ have also been investigated, often emerging as a natural extension within the same theoretical frameworks~\cite{Karliner:2014gca,Flynn:2003vz,Shah:2016vmd}.
The ${\rm \Omega}_{3c}$ baryon, in particular, predicted to have a mass around $4.8\,\text{GeV}$~\cite{Shah:2017jkr}, is expected to be stable against strong decays and to undergo weak decays with a lifetime comparable to that of singly charmed baryons~\cite{Bagan:1994dy}. 
Its experimental detection remains elusive, mainly due to the high production threshold and the challenges associated with reconstructing its multibody decay topologies in hadronic environments~\cite{Chang:2006eu,GomshiNobary:2005ur}. 
Fragmentation-based predictions for triply heavy baryons suggest extremely suppressed production rates at current colliders, with fragmentation probabilities in the $[10^{-5} \div 10^{-7}]$ range and total cross sections in the nanobarn regime, significantly complicating the direct observation of ${\rm \Omega}_{3c}$ states~\cite{Chen:2011mb,GomshiNobary:2005ur}.
Although direct observation of triply heavy baryons remains challenging, their presence might be inferred in decay products of exotic multiquark states, as suggested by theoretical predictions~\cite{Gershtein:2000nx} and recent experimental signals of pentaquarks with open charm and strangeness~\cite{LHCb:2015yax,LHCb:2019kea,LHCb:2020jpq}.

Triply heavy baryons also play a crucial role in the broader program of heavy-hadron spectroscopy and the classification of bound states in QCD~\cite{Brambilla:2019esw,Esposito:2016noz,Lebed:2016hpi,Faustov:2021qqf}. 
Their clean internal structure, devoid of light-quark effects, makes them ideal candidates for testing potential models, effective theories, and predictions from lattice QCD~\cite{Mathur:2018rwu,Padmanath:2013zfa,Francis:2018jyb}. 
Comparisons between triply heavy states and quarkonia, especially with regard to radial excitations and spin splittings, can help refine parameters such as the heavy-quark potential, effective coupling, and confining scale~\cite{Eichten:1994gt,Godfrey:1985xj}. 
In particular, analogies with fully heavy tetraquarks $|Q\bar{Q}Q\bar{Q}\rangle$ and pentaquarks $|QQQ\bar{Q}Q\rangle$~\cite{Ali:2017jda,LHCb:2020bwg,Karliner:2020vsi} suggest that ${\rm \Omega}_{3c}$ could be viewed as the baryonic anchor point of a larger spectrum of heavy-flavor objects, enabling unified treatments of their production through parton fragmentation. 
Such analogies also emerge in effective field theory descriptions, where nonrelativistic approximations lead to compact color configurations analogous to hydrogen- or heliumlike systems in atomic physics~\cite{Pineda:2011dg,Celiberto:2024mab,Celiberto:2024beg,Celiberto:2025dfe,Celiberto:2025ziy}.

Triply heavy baryons composed of charm and bottom quarks are particularly valuable for disentangling the interplay of fundamental forces at different distance scales in nuclear systems. Because of their Coulomb-like structure, the ground-state radii of these baryons scale inversely with the product of the quark mass and the strong coupling, making them significantly more compact than ordinary baryons. As a result, meson- and color-exchange dynamics differ substantially from those in conventional nuclear matter. 
Several theoretical predictions for their mass spectra exist, ranging from nonrelativistic variational estimates and potential model analyses~\cite{Llanes-Estrada:2011gwu,Wei:2016jyk,Yang:2019lsg,Gomez-Rocha:2023jfr,Najjar:2024deh} to recent quantum computing simulations based on the Cornell potential~\cite{deArenaza:2024dhe}. 
In particular, the triply charmed ${\rm \Omega}_{3c}$ baryon is expected to lie well within the discovery reach of future hadron colliders such as the High-Luminosity Large Hadron Collider (HL-LHC)~\cite{Apollinari:2015wtw}, with even greater prospects foreseen at the Future Circular Collider (FCC)~\cite{FCC:2025lpp,FCC:2025uan,FCC:2025jtd}, and may also be accessible at present and future lepton facilities like Belle~II~\cite{Belle-II:2010dht}.

The ATLAS, CMS, and LHCb experiments have already accumulated substantial datasets at 13 and 13.6~TeV, and projections for HL-LHC indicate that searches for triply heavy baryons could be feasible, especially in boosted regimes where modern tracking and vertexing can be exploited~\cite{ATLAS:2016bek,CMS:2024aqx,LHCb:2020frr}. 
Forward coverage, low transverse-momentum capabilities, and excellent mass resolution are key ingredients that make LHCb particularly well suited for this task. 
The proposed FCC, with its 100~TeV energy and unprecedented luminosity, would enhance discovery prospects even further, offering a unique opportunity to explore the heavy baryon sector beyond current limits.

From a phenomenological standpoint, triply heavy baryons are particularly compelling because of the unique nature of their formation mechanism.
More generally, the production of multiply heavy hadrons is expected to proceed predominantly through the fragmentation of high-energy partons into baryonic final states (see, \emph{e.g.}, Refs.~\cite{Braaten:1994bz,Braaten:1993rw,Braaten:1993mp,Braaten:1994xb,Braaten:1993jn,Kiselev:1994pu}).
This process is sensitive to both perturbative short-distance dynamics and nonperturbative aspects of hadronization.
In the case of heavy baryons, a practical framework for modeling this mechanism is offered by the quark-diquark picture~\cite{Anselmino:1992vg,Ebert:1995fp}, where a tightly bound diquark (\emph{e.g.}, $|cc\rangle$) forms first and subsequently hadronizes into the full baryon via the capture of an additional heavy quark~\cite{MoosaviNejad:2016qdx,Chang:2006eu}.

This diquark-based factorization simplifies the computation of fragmentation functions (FFs), which encode the probability density for a parton to hadronize into a specific final-state baryon carrying a given fraction of its momentum.
Since heavy-quark masses lie above the perturbative QCD threshold, the initial-scale inputs of FFs for heavy-flavored baryons incorporate both perturbative and nonperturbative components.
These can be modeled through effective parametrizations or derived from wave function overlaps inspired by potential models and nonrelativistic effective frameworks~\cite{Caswell:1985ui,Bodwin:1994jh,Braaten:1993rw,Cho:1995vh,Cho:1995ce,Bodwin:2005hm}.

Theoretical studies of fragmentation into triply heavy baryons have been carried out at leading order (LO) and next-to-leading order (NLO) in $\alpha_s$~\cite{Adamov:1997yk,Yang:2002gh,GomshiNobary:2004mq,MoosaviNejad:2017bda,MoosaviNejad:2017rvi,Delpasand:2019xpk}, with particular emphasis on charm- and gluon-initiated channels.
These analyses underscore the role of diquark correlations, binding energy corrections, and the underlying color structure in the baryon formation process.
To obtain realistic collider predictions, it is essential to incorporate higher-order corrections and parton evolution effects~\cite{Cacciari:1993mq,Buza:1996wv,Cacciari:2001cw,Mitov:2006wy}.

Experimentally, while no triply heavy baryon has been confirmed, their indirect detection might occur through decay chains of exotic hadrons~\cite{An:2019idk,Ortiz-Pacheco:2023kjn,Liu:2024mwn}. 
Multiquark states such as tetraquarks and pentaquarks with hidden or open heavy flavor (see Refs.~\cite{Chen:2016qju,Esposito:2016noz,Lebed:2016hpi} for a review) are routinely produced in LHC collisions, and their decay topologies may involve final-state three valence heavy quarks. 
In fact, ${\rm \Omega}_{3c}$ can emerge from hadronic transitions of heavy exotic hadrons or as a daughter particle in sequential decays~\cite{Chen:2011mb,Wang:2018utj,Yang:2019lsg}.

The motivation to construct realistic FFs for triply heavy baryons is thus multifold: improving our theoretical control on hadronization in the heavy sector, enabling predictions for rare baryon production at the LHC and future colliders, and contributing to the broader understanding of exotic matter in QCD, particularly through the role of ${\rm \Omega}_{3c}$ states as possible decay products of multiquark exotics such as pentaquarks.
To this end, we introduce in this work the {\tt OMG3Q1.0} sets~\cite{Celiberto:2025_OMG3Q10}, the first public release of collinear FFs for the ${\rm \Omega}_{3c}$ baryon. 
This set extends and complements recent efforts to characterize rare and exotic hadron production via fragmentation, such as those for doubly and fully heavy tetraquarks~\cite{Celiberto:2024_TQ4Q11,Celiberto:2024_TQHL11,Celiberto:2025_TQ4Q11_AVT,Celiberto:2024beg,Celiberto:2025dfe,Celiberto:2025ziy} and fully heavy pentaquarks~\cite{Celiberto:2025ipt}.

The {\tt OMG3Q1.0} functions are built by combining
diquarklike proxy model NLO inputs for both the charm-quark~\cite{MoosaviNejad:2017rvi} and gluon~\cite{Delpasand:2019xpk} channels at the lowest factorization scale.
Energy evolution to higher scales is performed using (Dokshitzer-Gribov-Lipatov-Altarelli-Parisi) DGLAP equations in a variable-flavor-number scheme (VFNS)~\cite{Mele:1990cw,Cacciari:1993mq,Buza:1996wv}, implemented via the Heavy-Flavor NonRelativistic evolution ({\HFNRevo}) scheme~\cite{Celiberto:2024mex,Celiberto:2024bxu,Celiberto:2024rxa}. 
Then, the evolved FFs are provided in \texttt{LHAPDF6} format \cite{Buckley:2014ana}, allowing for direct inclusion in high-energy simulation tools and cross section computations. 
They can be interfaced with resummation frameworks and parton-level Monte Carlo codes to generate realistic predictions for ${\rm \Omega}_{3c}$ production.

As a phenomenological application, we investigate the production of ${\rm \Omega}_{3c}$ plus jet systems at center-of-mass energies relevant for the (HL-)LHC and FCC.
The semi-inclusive process [${\rm p} {\rm p} \to {\rm \Omega}_{3c} + {\cal X} + {\rm jet}$] (see Fig.~\ref{fig:pictorial}) is studied in a $\NLLp$ hybrid framework that consistently merges NLO col\-linear factorization with the high-energy resummation of next-to-leading logarithmic (NLL$^+$) energy contributions and beyond.
Observables such as rapidity separations and transverse-momentum spectra are computed via the {\Jethad} code and the {\symJethad} symbolic plug-in~\cite{Celiberto:2020wpk,Celiberto:2022rfj,Celiberto:2023fzz,Celiberto:2024mrq,Celiberto:2024swu}.

This work contributes to the broader goal of developing a systematic and evolution-consistent description of the rare $\rm \Omega$ sector via VFNS fragmentation.
It offers a benchmark for future studies of triply heavy states and lays the groundwork for comparisons with lattice QCD, potential models, and heavy-flavor QCD phenomenology.

The remainder of this paper is organized as follows.
In Section~\ref{sec:FFs_heavy_flavor}, we briefly discuss general features of heavy-flavor fragmentation and then focus on the production of rare $\rm \Omega$ baryons.
Section~\ref{sec:FFs_architecture} details the architecture of our {\tt OMG3Q1.0} sets.
The hybrid factorization framework is outlined in Section~\ref{sec:hybrid_factorization}, and predictions for the HL-LHC and FCC are presented and discussed in Section~\ref{sec:results}.
Conclusions and outlook are given in Section~\ref{sec:conclusions}.

\section{Heavy-flavor fragmentation and the $\rm \Omega$ sector}
\label{sec:FFs_heavy_flavor}

In the first part of this section, we provide a concise overview of key aspects of heavy-flavor fragmentation, covering heavy-light hadrons, quarkonia, and exotic bound states (Section~\ref{ssec:FFs_highlights}).
We then discuss the main features of the diquarklike proxy model applied to triply heavy baryons (Section~\ref{ssec:FFs_diquark}).
Finally, we present the general structure of ${\rm \Omega}_{3c}$ initial-scale FFs (Section~\ref{ssec:FFs_O3c_general}).

\subsection{Highlights on heavy-flavor fragmentation}
\label{ssec:FFs_highlights}

The fragmentation of heavy-flavored hadrons is inherently more intricate than that of light hadrons. 
This complexity arises because heavy-quark masses in their lowest Fock states fall within the perturbative QCD domain. 
As a result, while light-hadron FFs encode only nonperturbative dynamics at the initial energy scale, heavy-hadron FFs require a combination of perturbative and nonperturbative elements.

For singly heavy hadrons such as $D$ mesons, $B$ mesons, and ${\rm \Lambda}_Q$ baryons, the fragmentation process unfolds in two distinct stages~\cite{Cacciari:1996wr,Cacciari:1993mq,Jaffe:1993ie,Kniehl:2005mk,Helenius:2018uul,Helenius:2023wkn}.  
The first stage involves a parton $i$, produced in a hard-scattering event with large transverse momentum, fragmenting into a heavy quark $Q$.  
Since $\alpha_s(m_Q) < 1$, this step can be computed within perturbative QCD.

This contribution, known as the short-distance coefficient (SDC) for the $[i \to Q]$ fragmentation process, occurs on a much shorter timescale than hadronization.  
The first NLO determination of SDCs for singly heavy hadrons was presented in Refs.~\cite{Mele:1990yq,Mele:1990cw}, with subsequent next-to-NLO refinements provided in Refs.~\cite{Rijken:1996vr,Mitov:2006wy,Blumlein:2006rr,Melnikov:2004bm,Mitov:2004du,Biello:2024zti}.

At later timescales, the heavy quark $Q$ transitions into a bound hadronic state.  
This second stage of fragmentation is entirely nonperturbative and is commonly modeled through phenomenological approaches~\cite{Kartvelishvili:1977pi,Bowler:1981sb,Peterson:1982ak,Andersson:1983jt,Collins:1984ms,Colangelo:1992kh} or effective field theories~\cite{Georgi:1990um,Eichten:1989zv,Grinstein:1992ss,Neubert:1993mb,Jaffe:1993ie}.

In mathematical terms, for a parton $i$ fragmenting into a singly heavy hadron $\HQ$ at the initial scale $\mu_{F,0}$ of the order of the mass of the heavy quark, $m_Q$, one has~\cite{Cacciari:1996wr,Cacciari_1997}
\begin{equation}
\label{FFs_HF_initial}
 D_{i}^{\HQ} (z, \mu_{F,0}) = 
 \int_z^1 \frac{\drv \xi}{\xi} D_i^Q (\xi, \mu_{F,0}) \, D_{\rm [np]}^{\HQ} \left( \frac{z}{\xi} \right) \;.
\end{equation}
Here, $D_i^Q$ is the perturbative part of the initial-scale FF, namely the SDC for an outgoing (massless) parton to fragment, via a perturbative QCD cascade, into the constituent (massive) heavy quark.
Conversely, $D_{\rm [np]}^{\HQ}$ depicts the nonperturbative part of the fragmentation, which is taken to be universal, independent of the generating parton $i$. 
It also does not depend on $\mu_{F,0}$.
The variable $z$ denotes the fraction of longitudinal momentum carried by the heavy hadron with respect to the fragmenting parton, while $\xi$ denotes the intermediate fraction passed to the constituent heavy quark before hadronization.

To construct a fully evolved set of FFs within the VFNS, it is crucial to account for energy evolution effects.  
Starting from initial-scale inputs, assumed to be free from scaling violations, the functions are evolved numerically using DGLAP timelike evolution equations at the required level of perturbative accuracy.

The two-step initial-scale fragmentation framework, originally devised for singly heavy hadrons, can be extended to quarkonia. 
In this case, the simultaneous presence of a heavy quark $Q$ and its antiquark $\bar{Q}$ within the lowest Fock state $|Q\bar{Q}\rangle$ adds an additional layer of complexity to the fragmentation description.  
The modern theoretical treatment of quarkonium production is based on nonrelativistic QCD (NRQCD)~\cite{Caswell:1985ui,Thacker:1990bm,Bodwin:1994jh,Cho:1995vh,Cho:1995ce,Leibovich:1996pa,Bodwin:2005hm} (for a pedagogical introduction, see Refs.~\cite{Grinstein:1998xb,Kramer:2001hh,QuarkoniumWorkingGroup:2004kpm,Lansberg:2005aw,Lansberg:2019adr}).  

NRQCD treats heavy-quark and antiquark fields as nonrelativistic degrees of freedom, enabling a systematic factorization between SDCs, which govern the perturbative production of the $[Q\bar{Q}]$ pair, and long-distance matrix elements (LDMEs), which encapsulate the nonperturbative hadronization dynamics.  
A physical quarkonium state is then expressed as a linear superposition of all possible Fock states, organized through a systematic expansion in both the strong coupling $\alpha_s$ and the relative velocity $v_{\cal Q}$ of the heavy-quark/antiquark pair.  

A key feature of NRQCD is its ability to describe quarkonium production mechanisms across a wide range of transverse momenta.  
At low transverse momentum, $q_T$, the dominant process involves the \emph{short-distance} formation of the $[Q\bar{Q}]$ pair in the hard scattering, followed by its nonperturbative evolution into a physical bound state.  
At higher $q_T$, however, an alternative mechanism---where a \emph{single parton} fragments into the quarkonium state plus additional radiation---begins to compete with, and eventually surpasses, the short-distance process.  

From a theoretical perspective, short-distance quark\-onium production can be interpreted as a fixed-flavor number scheme (FFNS)~\cite{Alekhin:2009ni} two-parton fragmentation, capturing higher-twist power corrections (see Refs.~\cite{Fleming:2012wy,Kang:2014tta,Echevarria:2019ynx,Boer:2023zit,Celiberto:2024mex,Celiberto:2024bxu,Celiberto:2024rxa} for more details).  
By contrast, the single-parton mechanism at large $q_T$ is a collinear VFNS fragmentation, its energy evolution being governed by the DGLAP equations.

The first LO calculations of the initial-scale inputs for both gluon and heavy-quark FFs of $S$-wave color-singlet vector quarkonia were performed in the early 1990s~\cite{Braaten:1993rw,Braaten:1993mp}.  
However, corresponding NLO refinements became available only recently~\cite{Zheng:2019gnb,Zheng:2021sdo}.  

Building upon these inputs, a pioneering set of VFNS, DGLAP-evolving FFs for vector quarkonia, {\tt ZCW19$^+$}, was introduced in Refs.~\cite{Celiberto:2022dyf,Celiberto:2023fzz}.  
Shortly thereafter, the {\tt ZCFW22} extension was developed to incorporate $B_c$ mesons~\cite{Celiberto:2022keu,Celiberto:2024omj}.  
Here, high-energy resummed production rates of charmed $B$ mesons, derived using the {\tt ZCFW22} FF framework, reinforced the experimental observation by the LHCb Collaboration~\cite{LHCb:2014iah,LHCb:2016qpe,Celiberto:2024omj} that the relative production rate of $\BCs$ mesons compared to singly bottomed $B$ mesons remains below 0.1\%~\cite{Celiberto:2024omj}.  
This result simultaneously
benchmarked for both the high-energy resummation framework and the NRQCD fragmentation applied to $B_c$ mesons.

Remarkably, recent studies indicate that NRQCD factorization can be suitably extended to explore the production mechanism of di-$\Jpsi$ excitations~\cite{LHCb:2020bwg,ATLAS:2023bft,CMS:2023owd}, interpreting these resonances as fully charmed, exotic tetraquarks~\cite{Zhang:2020hoh,Zhu:2020xni}.
From a NRQCD viewpoint, the production of a fully heavy $\TQQ$ tetraquark originates from the generation of a heavy-quark pair and its corresponding antiquark pair at short distances, with a characteristic energy scale set by the inverse of the heavy-quark mass.  
As with singly heavy hadrons and quarkonia, asymptotic freedom allows for the usual heavy-flavor fragmentation mechanism in two steps: an initial perturbative stage followed by a nonperturbative hadronization phase.  

The first NRQCD calculation of the initial-scale FF for the $[g \to \TQQ]$ $S$-wave channel in a color-singlet configuration appeared in Ref.~\cite{Feng:2020riv}.  
Building on this foundation, the recent {\tt TQ4Q1.0} VFNS sets combined this NRQCD-based gluon channel with an initial-scale input for the $[Q \to \TQQ]$ function, derived by adapting a Suzuki-model approach~\cite{Suzuki:1977km,Suzuki:1985up,Amiri:1986zv,Nejad:2021mmp}.  

Expanding this methodology, the first VFNS FFs for heavy-light tetraquarks, named {\tt TQHL1.0} determinations, were derived in Refs.~\cite{Celiberto:2023rzw,Celiberto:2024mrq}.  
The latest release, the {\tt TQ4Q1.1} and {\tt TQHL1.1} families~\cite{Celiberto:2024beg,Celiberto:2025dfe,Celiberto:2025ziy}, further refined the description by incorporating NRQCD-based modeling of also the $[Q \to \TQQ]$ initial-scale FF~\cite{Bai:2024ezn}, and improving the treatment of $\XQq$ fragmentation.

Finally, Ref.~\cite{Celiberto:2025ipt} accompanies and presents the first release of collinear FFs for fully charmed $|ccc\bar{c}c\rangle$ pentaquarks.
The resulting set {\tt PQ5Q1.0} embodies a consistent heavy-quark threshold DGLAP evolution from two possible initial-scale inputs for the heavy-quark channel: a compact direct multicharm state~\cite{Farashaeian:2024son} or a dicharm-charm-dicharm configuration~\cite{Farashaeian:2024cpd}.

\subsection{The diquarklike proxy model}
\label{ssec:FFs_diquark}

The quark-diquark model provides a simplified, yet powerful framework for describing the internal structure of baryons, wherein two quarks are treated as a tightly bound subsystem, referred to as a diquark, that interacts with the third quark~\cite{Gell-Mann:1964ewy}.  
This approach has been widely used in hadron spectroscopy and in the analysis of production and decay mechanisms of baryons, particularly those containing heavy quarks (see Refs.~\cite{Maiani:2004vq,Jaffe:2003sg,Guo:2013xga,DeSanctis:2016zph} for related discussions).  
Within this picture, diquarks are modeled either as scalar objects with spin-0 or as axial-vector states with spin-1.  
Their internal structure, inherently nonperturbative, is typically encoded through phenomenological form factors.

Scalar diquarks involve a single form factor and are generally associated with simpler spin configurations, while axial-vector diquarks require multiple form factors to describe their richer internal dynamics.  
Both types have been utilized in the modeling of spin-dependent processes in baryon formation, and also in the parametrization of polarized quark and gluon distributions inside the nucleon, especially within spectator models~\cite{Bacchetta:2008af,Bacchetta:2010si,Bacchetta:2020vty,Bacchetta:2024fci,Chakrabarti:2023djs,Banu:2024ywv}.

Early implementations of the quark-diquark framework to describe the fragmentation of light and heavy baryons can be found in Refs.~\cite{Nzar:1995wb,Ma:2001ri,Yang:2002gh} and~\cite{Falk:1993gb,Adamov:1997yk,MoosaviNejad:2017rvi,Delpasand:2019xpk}, respectively.  
The same formalism has also been extended to the analysis of exotic hadrons, such as pentaquarks~\cite{Maiani:2015vwa}, and to the mass spectroscopy of doubly and fully heavy tetraquarks using relativistic quasipotential approaches based on diquark-antidiquark interactions~\cite{Faustov:2020qfm,Faustov:2021hjs,Faustov:2022mvs}.

In the present work, we adopt a modeling strategy that has become standard in the literature~\cite{Adamov:1997yk,Martynenko:1996bt,MoosaviNejad:2017rvi,Delpasand:2019xpk}, in which the diquark is considered to be a scalar constituent.  
This choice is primarily motivated by computational convenience. 
It simplifies the spin algebra, reduces the number of independent form factors, and results in a more manageable analytical structure. 
It has been widely used in the modeling of unpolarized $J = 1/2$ baryons in different approaches.  
For example, in~\cite{Adamov:1997yk}, the scalar diquark model is used to calculate collinear FFs within a perturbative QCD framework, emphasizing its tractability.  
Similarly, the works in Refs.~\cite{Martynenko:1996bt,GomshiNobary:2007xk} show that adopting a scalar diquark allows one to construct models that are consistent with available data, while keeping the formalism analytically viable.

In contrast, the use of axial-vector diquarks introduces greater theoretical complexity, including the need for additional form factors and a more elaborate treatment of spin correlations, which limits their use in practical calculations.  
Nevertheless, axial-vector diquarks are essential for describing baryons with spin-$3/2$ and for incorporating polarization effects in both spin-$1/2$ and spin-$3/2$ states.  
In particular, any realistic modeling of the polarized fragmentation or production of the $\rm \Omega_{3c}^*$ resonance, with $J = 3/2$, requires a vector diquark configuration.

The scalar diquark approximation employed in this study should therefore be interpreted as an effective model for initiating a perturbative QCD-based description of fragmentation into unpolarized triply heavy baryons such as ${\rm \Omega}_{3c}$ states.  
More sophisticated treatments incorporating spin-1 diquarks could be pursued in future work to provide a more complete account of polarization and spin structure.

Interestingly, our adoption of the scalar diquark picture in the description of unpolarized ${\rm \Omega}_{3c}$ and ${\rm \Omega}_{3b}$ production is fully consistent with the color and spin symmetry requirements of the baryonic wave function, provided that the process is described within the two-step fragmentation mechanism.
In this framework, the heavy quark first fragments into a color-antitriplet heavy diquark, which subsequently hadronizes into the triply heavy baryon through nonperturbative QCD effects.

Although scalar diquarks alone do not manifest the full symmetrization structure expected for a $J=1/2$ baryon composed of three identical fermions, this structure can be effectively recovered through hadronization effects encoded in the purely nonperturbative component of the FF.\footnote{This separation between perturbative and nonperturbative structures is conceptually similar to the NRQCD-based description of quarkonium, where SDCs produce $[Q\bar{Q}]$ pairs in color-octet configurations, and the projection onto physical color-singlet states is performed via LDMEs.}

This interpretation, while not explicitly stated in~\cite{MoosaviNejad:2017rvi,Delpasand:2020mqv}, is in line with the modeling strategies adopted therein, where scalar diquarks are used in the indirect channel to compute FFs into ${\rm \Omega}_{3c}$ and ${\rm \Omega}_{3b}$, yielding consistent quantum number assignments and phenomenologically viable results.

\subsection{General structure of ${\rm \Omega}_{3c}$ FFs}
\label{ssec:FFs_O3c_general}

\begin{figure*}[!t]
\centering
\includegraphics[width=0.475\textwidth]{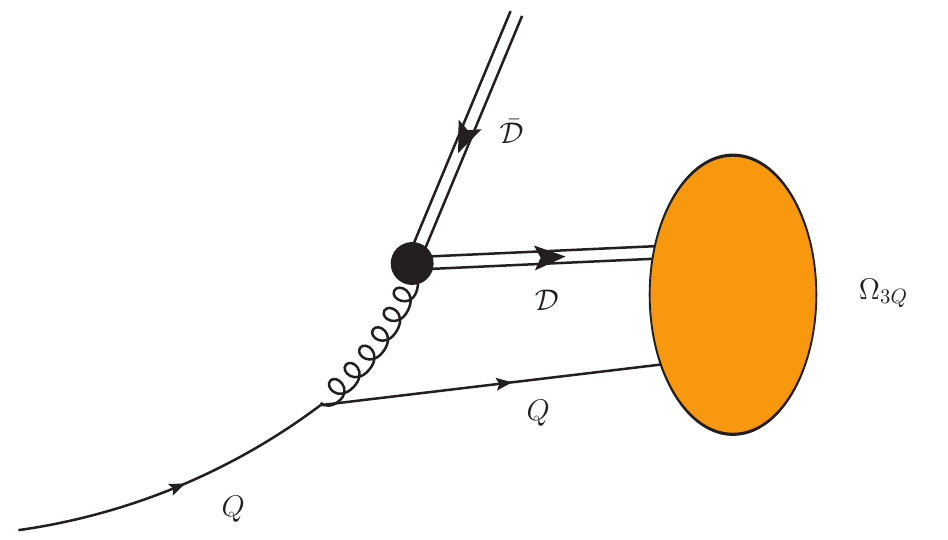}
\hspace{0.40cm}
\includegraphics[width=0.475\textwidth]{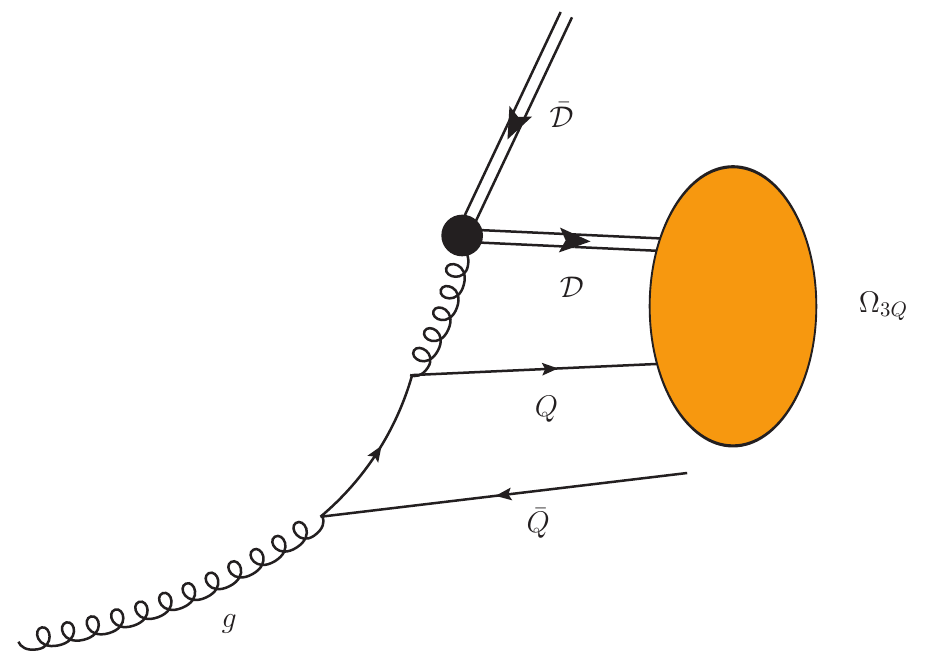}

\caption{Representative leading diagrams for the diquarklike proxy model of the initial-scale collinear fragmentation of a constituent heavy quark (left) or a gluon (right) into a color-singlet $S$-wave ${\rm \Omega}_{3Q}$ baryon.
Double lines stand for ${\cal D}$ or $\bar{{\cal D}}$ diquark states, while orange blobs represent the nonperturbative hadronization component of corresponding FFs.
Black bullet vertices denote scalar diquark form-factor couplings.}
\label{fig:O3Q_FF_diagrams}
\end{figure*}

The FF initial-scale inputs used in this study are extracted from existing calculations in the literature, which focus on the production of triply heavy baryons in a color-singlet configuration. 
Specifically, we rely on perturbative results for the quark-induced channel computed at LO in Ref.~\cite{MoosaviNejad:2017bda} and at NLO in Ref.~\cite{MoosaviNejad:2017rvi}, while for the gluon-induced channel we use the NLO result from Ref.~\cite{Delpasand:2019xpk}.
In all cases, the baryon is modeled as a color-singlet state formed from three charms, and the diquark is treated as a spin-0 scalar constituent.

Two main production scenarios are typically considered for phenomenological explorations. 
The first is the \emph{direct} fragmentation mechanism, in which the heavy quark fragments directly into the three-quark baryon without invoking any internal substructure. 
This approach involves three-body diagrams and includes the direct interaction among the three constituent quarks. 
The second scenario, which we adopt here, is the \emph{diquark} fragmentation model. 
In this framework, two of the three quarks are assumed to form a tightly bound scalar diquark, which then combines with the remaining quark to form the baryon. 
As mentioned before, this approximation significantly reduces the complexity of the calculation and has been widely used to describe the unpolarized production of $J = 1/2$ heavy baryons.

The so-called \emph{indirect} mechanism, as referred to in Refs.~\cite{MoosaviNejad:2017rvi,Delpasand:2019xpk}, corresponds precisely to the \emph{two-stage} fragmentation framework discussed in Section~\ref{ssec:FFs_highlights}, adapted to the diquark picture. 
From a theoretical standpoint, this structure offers a robust and systematically extensible framework for modeling the fragmentation of heavy-flavored hadrons.
First, it separates the perturbative production of an intermediate colored object (the diquark) from its nonperturbative hadronization into the physical baryon. 
Second, both the quark and gluon FFs have been calculated at NLO within this setup, making it a consistent and reliable foundation for a VFNS analysis.

The Suzuki model~\cite{Suzuki:1985up} is among the ingredients of the framework adopted in Refs.~\cite{MoosaviNejad:2017rvi,Delpasand:2019xpk}.
In that context, it plays a role analogous to that of NRQCD in quarkonium production.
Originally developed to describe the fragmentation of heavy quarks into baryons or mesons, the Suzuki picture provides both a structural and a computational foundation for the two-stage fragmentation process. 
It defines a perturbative FF transition amplitude, where the constituent heavy quark fragments into a constituent diquark and a spectator, followed by the hadronization of the diquark into a baryonic bound state. 
The model incorporates kinematic simplifications, such as the collinearity of final-state momenta, and introduces nonperturbative elements through the wave function at the origin and an effective diquark form factor.

The Suzuki model is used in both direct and diquark fragmentation scenarios studied in the literature, although its role is far more central in the latter. 
In the diquark-based case, it provides the full structure of the two-body fragmentation mechanism and enables the derivation of compact analytic expressions for the SDCs. 
This is particularly useful for modeling ${\rm \Omega}_{3c}$ baryons, where handling full QCD three-body fragmentation becomes analytically and symbolically prohibitive.

Adopting the Suzuki model to compute the perturbative part of the FF is motivated by both analytical practicality and physical consistency.
While rooted in perturbative QCD, the approach forgoes operator-based expansions in favor of a more ``diagrammatic’’ strategy.
This yields a simplified yet effective tool for evaluating SDCs, especially in the case of complex multiquark final states such as triply heavy baryons.

First, the Suzuki picture offers an analytically accessible framework that enables the closure of phase-space integrals and yields manageable expressions for amplitudes. 
Spin and color algebra are simplified, and the use of the wave function at the origin effectively absorbs the bound-state dynamics without invoking full nonperturbative QCD machinery. 
This is especially valuable in contexts where rigorous QCD calculations requiring renormalization, operator regularization, or twist expansions are unfeasible or unavailable.

Second, the Suzuki model retains an explicit connection to the internal structure of the hadron. Rather than treating the baryon as a pointlike final state, it describes the fragmentation process using constituent-level vertices, including spin-dependent terms, propagators for internal degrees of freedom, and color couplings. 
This construction ensures that the hard subprocess remains sensitive to the underlying baryonic wave function, preserving the link between perturbative production and hadron-level observables.

Third, the model is particularly well suited for baryons like ${\rm \Omega}_{3c}$, where direct short-distance transitions in full perturbative QCD involve the algebraic disentangling of intertwined Dirac, Lorentz, and color structures.
By rephrasing the process as a two-body fragmentation, where the diquark is treated as an effective, quasielementary object, the Suzuki model circumvents these challenges while preserving the overall dynamical content.

The analogy with NRQCD is particularly instructive. 
Just as NRQCD allows for a systematic separation of short-distance and long-distance contributions in quarkonium production, the Suzuki model supports a clean factorization between perturbative dynamics, described by the SDCs, and nonperturbative hadronization. 
Both frameworks rely on the use of effective composite degrees of freedom (diquark or heavy-quark nonrelativistic fields), which serve to simplify the treatment of fragmentation processes while maintaining fidelity to the underlying physics.

In the specific implementation adopted in Refs.~\cite{MoosaviNejad:2017rvi,Delpasand:2019xpk}, the nonperturbative dynamics of the baryonic bound state are incorporated via a light-cone distribution amplitude, which effectively plays the role of a wave function at the origin. 
Following the approach of Lepage and Brodsky~\cite{Lepage:1980fj}, this distribution amplitude is approximated by a Dirac $\delta$ function in the momentum fraction variables, thereby forcing the constituents to move collinearly inside the baryon. 
This assumption leads to a compact analytic form for the fragmentation amplitude, where the nonperturbative input does not appear as a separate factor but is instead embedded within the perturbative expression itself. 
This modeling strategy has already been successfully employed in the study of heavy-quark fragmentation into fully heavy-light tetraquarks~\cite{Nejad:2021mmp,Celiberto:2023rzw} and pentaquark-charmonium states~\cite{Farashaeian:2024son,Farashaeian:2024cpd,Celiberto:2025ipt}, where a similar factorization between short- and long-distance dynamics is achieved through an effective wave function treatment.

A crucial component in the construction of our initial-scale FFs is the modeling of the nonperturbative transition from the constituent diquark ${\cal D}$ to the physical baryon ${\rm \Omega}_{3c}$. 
Following~\cite{MoosaviNejad:2017rvi,Delpasand:2019xpk}, in this work we adopt the phenomenological Peterson-Schlatter-Schmitt-Zerwas (PSSZ) parametrization~\cite{Peterson:1982ak}, which has been widely used in heavy-flavor fragmentation studies and is particularly suited for scenarios involving hadrons with multiple heavy quarks.
We have
\begin{equation}
\label{HFF_PSSZ}
 D_{\cal D}^{{\rm \Omega}_{3c}} (z) \equiv D_{\rm [np]}^{\rm [PSSZ]} (z) = {\cal N}_{\cal P} \frac{z(1-z)^2}{\left[(1-z)^2 + b_{\cal P} z \right]^2} \;,
\end{equation}
with the normalization factor
\begin{equation}
\begin{split}
\label{HFF_PSSZ_Nrm}
 {\cal N}_{\cal P} &\,=\, 
  \left\{
 \frac{b_{\cal P}^2-6b_{\cal P}+4}{(4-b_{\cal P})\sqrt{4b_{\cal P}-b_{\cal P}^2}}
  \right.
  \\ &\hspace{-0.20cm}
 \times \,
  \left[ 
 \arctan \frac{b_{\cal P}}{\sqrt{4 b_{\cal P} - b_{\cal P}^2}} + \arctan \frac{2-b_{\cal P}}{\sqrt{4 b_{\cal P} - b_{\cal P}^2}} 
  \right]
  \\ &\hspace{-0.20cm}
 + \,
  \left.
 \frac{1}{2} \ln b_{\cal P} + \frac{1}{4-b_{\cal P}} 
  \right\} \;.
\end{split}
\end{equation}
being obtained by summing over all hadrons containing the charm quark~\cite{Cacciari_1997} (see also Section~IV of Ref.~\cite{Peterson:1982ak}).
From the original derivation of the model, we note that the $b_{\cal P}$ cannot be calculated exactly, since it is intrinsically connected to nonperturbative effects.

Indeed, for a singly charmed hadron, say a $D$ meson with a leading Fock state $| c \bar{q} \rangle$, the value of $b_{\cal P}$ is related to the charm-quark mass by $b_{\cal P} \approx {\rm \Lambda}/m_c$, where ${\rm \Lambda}$ stands for a hadronic scale of the order of the mass of the constituent light antiquark, $m_{\bar q}$.
In such a case, one finds that $\langle z \rangle = 1 - \sqrt{b_{\cal P}}$, which is in line with the prediction of heavy-quark FFs scaling linearly with the heavy-quark mass~\cite{Suzuki:1977km,Bjorken:1977md,Kinoshita:1985mh,Cacciari:1996wr}.
On the contrary, for a fully charmed baryon ${\rm \Omega}_{3c}$, the $b_{\cal P}$ is connected to the charm and anticharm masses, which are above the perturbative threshold.
Following Refs.~\cite{Adamov:1997yk,MoosaviNejad:2017rvi,Delpasand:2019xpk}, we set $b_{\cal P} = [m_c/(m_c + m_{\bar{c}})]^2 = 1/4$.

This choice is supported by previous studies, where the Peterson form is explicitly used to model the nonperturbative stage of the diquark hadronization in fragmentation to triply heavy baryons~\cite{MoosaviNejad:2017rvi,Delpasand:2019xpk}. 
Work in Ref.~\cite{MoosaviNejad:2017rvi} clearly emphasizes that the Peterson function effectively captures the essential nonperturbative dynamics of the fragmentation process and is compatible with the overall two-step fragmentation scheme defined through the Suzuki model. 
Similarly, Ref.~\cite{Delpasand:2019xpk} supports the use of the Peterson parametrization on the basis of its widespread adoption and its ability to encode realistic hadronization patterns for heavy baryons. 
Moreover, the Peterson form complements the output of the perturbative calculation by providing a universal and physically interpretable nonperturbative component that fits naturally into collinear factorization frameworks.

The Peterson model, by contrast, was originally formulated to describe the fragmentation of heavy quarks in which the initiating heavy quark carries most of the hadron momentum. 
This structure is naturally suited to hadrons containing heavy constituents, since the resulting hadron tends to follow the direction and kinematics of the fragmenting parton. 
In the case of triply heavy baryons, where all valence quarks are heavy, the same reasoning applies: the baryon is formed predominantly from the massive fragments of the initial parton, and thus inherits a large fraction of its momentum.
In this respect, the Peterson function provides a realistic and physically consistent parametrization of the nonperturbative transition, capturing the expected peak at large z and the suppression at small z, in line with the kinematics of systems such as triply heavy baryons.

This behavior is particularly well suited for modeling the hadronization of a heavy diquark into a triply heavy baryon, where the recombination process is dominated by massive constituents and the baryon inherits a substantial fraction of the momentum of the initiating quark or gluon. 
The simple analytic structure of the model, governed by a single shape parameter $b_{\cal P}$, allows for the adjustment to match the internal mass hierarchy and spatial configuration of the final-state baryon, while ensuring smooth integration within collinear convolution schemes.

In this context, the Peterson form meets several important criteria: it is grounded in heavy-flavor dynamics, it offers analytical tractability, it aligns with the factorized structure emerging from the Suzuki picture, and it preserves consistency with the DGLAP evolution in the VFNS. 
For these reasons, we consider it a natural and robust choice for describing the nonperturbative input to the ${\rm \Omega}_{3c}$ initial-scale FF in our two-step formalism.

The compact expression for the FF of a parton $i$ generating a ${\rm \Omega}_{3Q}$ baryon at the initial scale $\mu_{F,0}$ reads
\begin{equation}
\label{FFs_O3c_initial}
 D_{i}^{{\rm \Omega}_{3c}} (z, \mu_{F,0}) = 
 \int_z^1 \frac{\drv \xi}{\xi} D_i^{\cal D} (\xi, \mu_{F,0}) \, D_{\cal D}^{{\rm \Omega}_{3c}} \left( \frac{z}{\xi} \right) \;.
\end{equation}
Equation~\eqref{FFs_O3c_initial} essentially matches the general structure of the heavy-hadron initial-scale FF as in Eq.~\eqref{FFs_HF_initial}, with two suitable adjustments.
We notice that
\begin{subequations}
\begin{align}
\label{FFs_O3c_vs_HF_initial_a}
 D_i^Q (\xi, \mu_{F,0})
  \quad &\longrightarrow \quad
 D_i^{\cal D} (\xi, \mu_{F,0})
  \;,
  \\[0.25cm]
\label{FFs_O3c_vs_HF_initial_b}
 D_{\rm [np]}^{\HQ} \left( \frac{z}{\xi} \right)
  \quad &\longrightarrow \quad
 D_{\cal D}^{{\rm \Omega}_{3c}} \left( \frac{z}{\xi} \right)
  \;.
\end{align}
\end{subequations}
Here we have SDCs describing the perturbative splitting of a massless parton $i$ into a diquark state ${\cal D}$ (Eq.~\eqref{FFs_O3c_vs_HF_initial_a}), followed by the long-distance, nonperturbative hadronization of the diquark into the observed triply heavy baryon ${\rm \Omega}_{3c}$ (Eq.~\eqref{FFs_O3c_vs_HF_initial_b}).
In other words, the diquark ${\cal D}$ replaces the $Q$ entering the lowest Fock state of the singly heavy hadron $\HQ$.
As mentioned, the nonperturbative $D_{\cal D}^{{\rm \Omega}_{3c}}$ will be modeled according to PSSZ as in Eq.~\eqref{HFF_PSSZ}, while gluon and charm SDCs will be presented and discussed in the next section.

\section{The {\tt OMG3Q1.0} architecture}
\label{sec:FFs_architecture}

In this section, we outline our scheme for developing the hadron-structure-oriented {\tt OMG3Q1.0} VFNS FF determinations for fully charmed ${\rm \Omega}$ baryons, starting from diquarklike proxy model inputs for both the charm-quark and gluon channels at the initial energy scales and accounting for a threshold-consistent DGLAP as follows from the {\HFNRevo} methodology~\cite{Celiberto:2024mex,Celiberto:2024bxu,Celiberto:2024rxa}.

All calculations necessary to build our {\tt OMG3Q1.0} set were carried out using {\symJethad}, the newly developed \textsc{Mathematica}~\cite{Mathematica_V14-2} plug-in of {\Jethad} \cite{Celiberto:2020wpk,Celiberto:2022rfj,Celiberto:2023fzz,Celiberto:2024mrq,Celiberto:2024swu}, designed for the symbolic manipulation of analytical formulas pertinent to the hadronic structure and precision QCD.

Sections~\ref{ssec:FFs_c} and~\ref{ssec:FFs_g} detail the charm and gluon inputs, respectively. 
The DGLAP/{\HFNRevo} timelike evolution of the {\tt OMG3Q1.0} functions is then analyzed and discussed in Section~\ref{ssec:FFs_OMG3Q10}.

\subsection{Initial-scale charm fragmentation}
\label{ssec:FFs_c}

The initial-scale FF for the $[c \to {\rm \Omega}_{3c}]$ transition in the diquarklike scenario is diagrammatically shown in the left panel of Fig.~\ref{fig:O3Q_FF_diagrams}.
In this representation, the double lines denote dicharm states, ${\cal D} \equiv |cc\rangle$, or their corresponding antidiquark partners, $\bar{{\cal D}} \equiv |\bar{c}\bar{c}\rangle$.
The orange blob illustrates the nonperturbative hadronization component of the FF, labeled as $D_{\rm [np]}^{\HQ} \equiv D_{\cal D}^{{\rm \Omega}_{3c}}$.
The black bullet vertex represents the scalar diquark form-factor coupling.

Although not shown explicitly in the formul{\ae} presented below, our calculation includes a scalar diquark form factor at the black bullet vertex displayed in the left panel of Fig.~\ref{fig:O3Q_FF_diagrams}. 
This factor models the internal structure of the diquark and is treated following the approach adopted in Ref.~\cite{MoosaviNejad:2017rvi}, where a simple pole behavior in transverse-momentum space is assumed, with a characteristic scale of approximately 5~GeV. 
We choose not to display it analytically to maintain a compact expression for the initial-scale FF, but its effect is fully accounted for in the diagrammatic formulation and implicitly absorbed into the normalization of the perturbative amplitude.

We note that the diagram on the left panel of Fig.~\ref{fig:O3Q_FF_diagrams} corresponds to the $[c \to (c , {\cal D}) + \bar{\cal D}]$ splitting.
By swapping all charm quarks with anticharms, one obtains the $[\bar{c} \to (\bar{c} , \bar{\cal D}) + {\cal D}]$ process.
A detailed analysis of potential asymmetries in the production of triply charmed $\rm \Omega$ baryons and antibaryons is left for future dedicated studies.
In this work, we assume full symmetry in the formation mechanisms of ${\rm \Omega}_{3c}$ and $\bar{{\rm \Omega}}_{3c}$ states.
Accordingly, our predictions are based on observables sensitive to the inclusive, charge-averaged emission of baryons and antibaryons.
Assuming this symmetry holds, the $c$ and $\bar{c}$ fragmentation channels are treated on equal footing (for comparison with the light-hadron case, see Ref.~\cite{Bertone:2018ecm}).

Using {\symJethad}~\cite{Celiberto:2020wpk,Celiberto:2022rfj,Celiberto:2023fzz,Celiberto:2024mrq,Celiberto:2024swu} and {\FeynCalc}~\cite{Mertig:1990an,Shtabovenko:2016sxi,Shtabovenko:2020gxv}, we symbolically calculated and obtained the explicit form of the LO and NLO initial-scale charm SDC, finding agreement with Refs.~\cite{MoosaviNejad:2017bda,MoosaviNejad:2017rvi}, respectively.
As for the LO case, we write
\begin{equation}
\begin{split}
\label{FFs_O3c_c}
 D_{c}^{\cal D} (z, \mu_{F,0}) \,&\equiv 
 \as^2
 \drv_{c}^{\rm [LO]} (z; \mu_{F,0}) \\[0.10cm]
 \,&+\,
 \as^3 \, \drv_{c}^{\rm [NLO]} (z; \mu_{F,0})
 \;.
\end{split}
\end{equation}
with $\as \equiv \as(5m_c)$ and
\begin{equation}
\label{FFs_O3c_c_d_LO}
 \drv_{c}^{\rm [LO]} (z; \mu_{F,0}) =
 {\cal N}_c(z)
 \frac{{\cal S}_c^{\rm [LO]}(z; {\cal R}_{q_T/c})}{{\cal T}_c^{\rm [LO]}(z; {\cal R}_{q_T/c})}
 \;,
\end{equation}
where
\begin{equation}
\label{R_qTc}
  {\cal R}_{q_T/c}^2 \equiv \vqTTa/m_c^2
 \;,
\end{equation}
\begin{equation}
\label{FFs_O3c_c_d_norm}
 {\cal N}_c(z) = \frac{\pi^2}{\sqrt{6}} \, f_{\cal B}^2 C_F^2 \, z^3(1-z)^3
 \;,
\end{equation}
\begin{equation}
\begin{split}
\label{FFs_O3c_Q_d_num_LO}
  {\cal S}_c^{\rm [LO]}&(z; {\cal R}_{q_T/c}) 
  \,=\, 
  4(41z^2 - 80z + 96 + 256/z^2) \\
  \,&+\,
  {\cal R}_{q_T/c}^2 (8z^3 + 5z^2 - 16z + 16) \\
  \,&+\,
  {\cal R}_{q_T/c}^4 \, 4z^2
  \;,
\end{split}
\end{equation}
and
\begin{equation}
\label{FFs_O3c_Q_d_den_LO}
\begin{split}
 {\cal T}_c^{\rm [LO]}&(z; {\cal R}_{q_T/c}) 
  \,=\, 
  ({\cal R}_{q_T/c}^2 + (4-3z)^2)^2 \\
  \,&\times\,
  {\cal R}_{q_T/c}^2 \, z^2 + z^2 - 16 z + 16)^2
 \;.
\end{split}
\end{equation}
The parameter $f_{\cal B}$ in Eq.~\eqref{FFs_O3c_c_d_norm} represents the decay constant of the triply heavy baryon.

We adopt the value $f_{\cal B} = 0.25$~GeV, in line with common assumptions in the literature~\cite{ParticleDataGroup:2020ssz}.
The NLO correction to the charm SDC was originally derived in Ref.~\cite{MoosaviNejad:2016qdx} for singly heavy mesons, and then employed for triply heavy baryons (see Ref.~\cite{MoosaviNejad:2017rvi}).
It reads
\begin{equation}
\label{FFs_O3c_c_d_NLO}
 \drv_{c}^{\rm [NLO]} (z; \mu_{F,0}) =
 {\cal N}_c(z)
 \frac{{\cal S}_c^{\rm [NLO]}(z; {\cal R}_{q_T/c})}{{\cal T}_c^{\rm [NLO]}(z; {\cal R}_{q_T/c})}
 \;,
\end{equation}
where
\begin{equation}
\begin{split}
\label{FFs_O3c_Q_d_num_NLO}
  {\cal S}_c^{\rm [NLO]}&(z; {\cal R}_{q_T/c}) 
  \,=\, 
  96 \pi z
\left[\right.
  {\cal R}_{q_T/c}^{14} \, (17-20 z) z^{14} \\
\,&+\,
  {\cal R}_{q_T/c}^{12} \, (240 z^4-1664 z^3 \\
\,&+\,
  3813 z^2-4130 z+1711) z^{12} \\
\,&+\,
  {\cal R}_{q_T/c}^{10} \, (4368 z^6-39072 z^5+140635 z^4 \\
\,&-\,
  269144 z^3+300490 z^2 \\
\,&-\,
  189616 z+52243) z^{10} \\
\,&+\,
  {\cal R}_{q_T/c}^8 \, (20976 z^8-208040 z^7 \\
\,&+\,
  870151 z^6-2070634 z^5+3255115 z^4 \\
\,&-\,
  3533740 z^3+2600201 z^2 \\
\,&-\,
  1202330 z+268205) z^8 \\
\,&+\,
  {\cal R}_{q_T/c}^6 \, 4 (1-z)^2 (11256 z^8-75035 z^7 \\
\,&-\,
  151779 z^6+2535886 z^5 \\
\,&-\,
  8913460 z^4+15864355 z^3 \\
\,&-\,
  15844619 z^2+8395110 z-1823490) z^6 \\
\,&+\,
  {\cal R}_{q_T/c}^4 \, 4 (1-z)^4 (12324 z^8-22830 z^7 \\
\,&-\,
  2045973 z^6+14641206 z^5-42277675 z^4 \\
\,&+\,
  65338590 z^3-57387186 z^2 \\
\,&+\,
  26330832 z-4933872) z^4 \\
  \,&+\,
  {\cal R}_{q_T/c}^2 \, 48 (1-z)^6 (563 z^8+2272 z^7 \\
\,&-\,
  252314 z^6+1752016 z^5-5068605 z^4 \\
\,&+\,
  7770780 z^3-6800544 z^2 \\
\,&+\,
  3505032 z-849528) z^2 \\
\,&+\,
  144 (1-z)^8 (41 z^8+438 z^7-36063 z^6 \\
\,&+\,
  258240 z^5-833328 z^4+1614600 z^3 \\
\,&-\,
  1793016 z^2+676512 z-198288)
\left.\right]
  \;,
\end{split}
\end{equation}
and
\begin{equation}
\label{FFs_O3c_Q_d_den_NLO}
\begin{split}
\hspace{-0.15cm}
 {\cal T}_c^{\rm [NLO]}(z; {\cal R}_{q_T/c}) 
  \,&=\, 
 [{\cal R}_{q_T/c}^2 \, z^2 + 4 (3-2 z)^2]^2 \\
\,&\times\,
 [{\cal R}_{q_T/c}^2 \, z^2 + (6-z)^2]^2  \\
\,&\times\,
[{\cal R}_{q_T/c}^2 \, z^2 + (1-z)^2] \\
\,&\times\, 
[{\cal R}_{q_T/c}^2 \, z^2 + 36(1-z)^2]^2  \\
\,&\times\,
[{\cal R}_{q_T/c}^2 \, z^2 + z^2-35 z+36]
  \;.
\end{split}
\end{equation}
Based on LO kinematics, we set the initial scale of charm fragmentation to $\mu_{F,0} \equiv 5 m_c$ (see Fig.~\ref{fig:O3Q_FF_diagrams}, left panel).

Our charm FF differs from the one originally introduced in Ref.~\cite{MoosaviNejad:2017rvi} in two main aspects.
First, in that study, the constant ${\cal N}_c(z)$ in Eq.~\eqref{FFs_O3c_c_d_norm} was not explicitly computed but was fixed by means of a normalization condition.
Second, the choice of the $\vqTTa$ parameter, directly connected with the use of the Suzuki picture, requires further refinement.

The Suzuki model effectively accounts for spin correlations and serves as a proxy for transverse-momentum-dependent (TMD) FFs.
In the collinear limit, instead of integrating over the squared transverse momentum of the outgoing charm quark, one may replace it with its average value, $\vqTTa$.
This substitution renders $\vqTTa$ a free parameter that must be fixed phenomenologically.
Increasing $\vqTTa$ shifts the FF peak to lower values of $z$ while reducing the overall normalization~\cite{GomshiNobary:1994eq}.

The initial-scale quark FF presented in Ref.~\cite{MoosaviNejad:2017rvi} was obtained by adopting $\vqTTa = 1\,\text{GeV}^2$, regarded as an upper-limit estimate for the average squared transverse momentum.
In this work, we introduce a more refined selection of $\vqTTa$, based on a balanced and phenomenologically motivated approach consistent with the exploratory goals of our analysis.
This improvement is based on a preliminary refinement introduced in our previous investigation of collinear fragmentation into fully charmed tetraquarks \cite{Celiberto:2024mab}.
There, we drew on phenomenological insights from fragmentation studies of various hadrons produced in proton collisions.
In particular, heavy-quark FFs for both light hadrons~\cite{Celiberto:2016hae,Celiberto:2017ptm,Bolognino:2018oth,Celiberto:2020wpk} and heavy-flavored ones~\cite{Celiberto:2021dzy,Celiberto:2021fdp,Celiberto:2022dyf,Celiberto:2022keu} were observed to be typically probed at average longitudinal-momentum fractions $\langle z \rangle > 0.4$.

Based on a numerical analysis, we determined that setting $\vqTTa_{\TQc} = 70\,\text{GeV}^2$ for charm-to-$\TQc$ FFs results in an average $z$ greater than 0.4 and preserves a comparable order of magnitude between the quark and gluon channels.
Extending this strategy to fully charmed $|ccc\bar{c}c\rangle$ pentaquarks, we adopted $\vqTTa_{\PQc} = 90\,\text{GeV}^2$ in Ref.~\cite{Celiberto:2025ipt}.
Here, we fix $\vqTTa_{{\rm \Omega}_{3c}}$ by correlating it with the peak position of the initial charm FFs.
Following the same methodology used for $\vqTTa_{\TQc}$ and $\vqTTa_{\PQc}$, we performed a numerical scan and selected $\vqTTa_{{\rm \Omega}_{3c}} = 60\,\text{GeV}^2$.

A deeper justification supports our choice, going beyond heuristic arguments.
Foundational studies on heavy-flavor fragmentation~\cite{Suzuki:1977km,Bjorken:1977md,Kinoshita:1985mh,Peterson:1982ak} have shown that heavy-quark FFs typically peak at large values of $z$, with binding effects scaling in proportion to the heavy-quark mass.
To illustrate this behavior, consider the fragmentation process leading to a $D$ meson, whose lowest Fock state is $|c{\bar q}\rangle$, with total momentum $k$ and mass $M_D$.

In this context, the constituent heavy quark and the light antiquark are assumed to move with approximately the same velocity, $v \equiv v_c \simeq v_q$.
Their momenta can thus be expressed as $k_c \equiv z k = m_c v$ for the charm quark and $k_q = {\rm \Lambda}_q v$ for the light antiquark, where ${\rm \Lambda}_q$ is a hadronic mass scale of order ${\rm \Lambda_{\rm QCD}}$.
Since the mass of the $D$ meson satisfies $M_D \approx m_c$, one finds $m_c v \approx k = k_c + k_q = z m_c v + {\rm \Lambda}_q v$.
This implies the approximate relation $\langle z \rangle_c \approx 1 - {\rm \Lambda}_q / m_c$, where the subscript `$c$’ indicates the $[c \to D]$ FF.

As pointed out in Refs.~\cite{Celiberto:2024mab,Celiberto:2025ipt}, this behavior does not generally apply to fully heavy-flavored states such as quarkonia, triply charmed baryons, tetracharms, or pentacharms.
In the valence configuration of these systems, no soft scale is present, as their lowest Fock components contain only heavy quarks.
For fully heavy rare baryons or exotics like $\TQc$ or $\PQc$ states, the complex dynamics among the three, four, or five heavy constituents hinders a simple kinematic determination of the peak position of the FF.

As a preliminary example, we present the $z$ dependence of the $[c \to {{\rm \Omega}_{3c}}]$ initial-scale FFs used in the {\tt OMG3Q1.0} set.
To estimate uncertainties at the starting scale, we apply a simplified, diagonal DGLAP evolution at NLO, similar to that in Ref.~\cite{Celiberto:2024mab}, but using only the $P_{qq}$ kernel for the charm FF.
The left panel of Fig.~\ref{fig:FFs_initial-scale} shows the $z$-weighted charm FF, with $\mu_{F,0} = 5 m_c$ and scale variations in the range $4m_c < \mu_F < 6m_c$.
The $z$-weighted FF peaks in the $0.45 < z < 0.5$ range and drops to zero at the end points, as expected.

The small negative dip visible near $z \gtrsim 0.8$ may appear anomalous at first glance.
However, dedicated numerical tests show that this feature originates from the choice of a relatively large value of $\vqTTa$, higher than the one adopted in Ref.~\cite{MoosaviNejad:2017rvi}.
This has no practical impact, as our tests confirm that FFs contribute dominantly to hadroproduction observables in the region $0.4 \lesssim z \lesssim 0.6$ (see Section~\ref{ssec:FFs_OMG3Q10}).
For $z > 0.8$, we have explicitly verified that no instability arises in high-energy cross section predictions.

\subsection{Initial-scale gluon fragmentation}
\label{ssec:FFs_g}

The initial-scale FF for the $[g \to {\rm \Omega}_{3c}]$ transition in the diquarklike scenario is diagrammatically shown in the right panel of Fig.~\ref{fig:O3Q_FF_diagrams}.
Although both the charm- and gluon-initiated channels are considered leading within their respective topologies, they differ in their perturbative order.
As shown in Fig.~\ref{fig:O3Q_FF_diagrams}, the gluon-induced process involves one additional QCD vertex compared to the charm-induced one, thus contributing at a higher order in the strong coupling expansion.
Specifically, the $[c \to {\rm \Omega}_{3c}]$ channel is known both at LO, corresponding to the left diagram, and at NLO, as discussed in Section~\ref{ssec:FFs_c}.
In contrast, the $[g \to {\rm \Omega}_{3c}]$ channel is currently known only at $\mathcal{O}(\alpha_s^3)$, which corresponds to the diagram shown in the right panel.

This difference comes from the fact that a gluon, which does not carry a charm quantum number, must generate at least three charm quarks to produce ${\rm \Omega}_{3c}$.
Therefore, the first perturbative order at which such a process can occur is $\mathcal{O}(\alpha_s^3)$, since the gluon must emit a $[c\bar{c}]$ pair to supply the free charm quark, generate a $|cc\rangle$ system to form the diquark (either directly or through an equivalent intermediate process) and include at least one additional QCD vertex to complete the baryon formation diagram.
For consistency with this perturbative hierarchy, we denote the initial-scale gluon FF as the NLO function, while reserving the LO label for the lowest-order charm FF.

With {\symJethad}~\cite{Celiberto:2020wpk,Celiberto:2022rfj,Celiberto:2023fzz,Celiberto:2024mrq,Celiberto:2024swu} and through the interface to {\FeynCalc}~\cite{Mertig:1990an,Shtabovenko:2016sxi,Shtabovenko:2020gxv}, we symbolically computed the explicit form of the NLO initial-scale gluon SDC, originally provided in Ref.~\cite{Delpasand:2019xpk}.
One has
\begin{equation}
\label{FFs_O3c_g}
 D_{g}^{\cal D} (z, \mu_{F,0}) \,\equiv\, 
 \as^3 \, \drv_{g}^{\rm [NLO]} (z; \mu_{F,0}) \;.
\end{equation}
with $\as \equiv \as(6m_c)$ and
\begin{equation}
\label{FFs_O3c_g_d}
 \drv_{g}^{\rm [NLO]} (z; \mu_{F,0}) =
 {\cal N}_g(z)
 \frac{{\cal S}_g^{\rm [NLO]}(z; {\cal R}_{q_T/c})}{{\cal T}_g^{\rm [NLO]}(z; {\cal R}_{q_T/c})}
 \;,
\end{equation}
where
\begin{equation}
\label{FFs_O3c_g_d_norm}
 {\cal N}_g(z) = \frac{2 \pi^3}{3} \, f_{\cal B}^2 C_F^2 \, z^3(1-z)^2
 \;,
\end{equation}
\begin{equation}
\begin{split}
\label{FFs_O3c_g_d_num_NLO}
  {\cal S}_g^{\rm [NLO]}(z; {\cal R}_{q_T/c}) 
  \,&=\, 
  8(16z^2-32z+15) \\
  \,&+\,
  {\cal R}_{q_T/c}^2 \, 2z^2(4z^2-20z+17) \\ 
  \,&+\,
  {\cal R}_{q_T/c}^4 \, z^4
\end{split}
\end{equation}
and
\begin{equation}
\label{FFs_O3c_g_d_den_NLO}
\begin{split}
 {\cal T}_g^{\rm [NLO]}&(z; {\cal R}_{q_T/c}) 
  \,=\, 
  (4 + z^2{\cal R}_{q_T/c}^2)^5
 \;.
\end{split}
\end{equation}
As in Section~\ref{ssec:FFs_c} (see Eq.~\eqref{R_qTc}), here we set ${\cal R}_{q_T/c}^2 \equiv \vqTTa/m_c^2$, with $\vqTTa \equiv \vqTTa_{{\rm \Omega}_{3c}} = 60\,\text{GeV}^2$.
In contrast to the charm case, the LO kinematics of gluon fragmentation set the initial scale at $\mu_{F,0} \equiv 6 m_c$ (see Fig.~\ref{fig:O3Q_FF_diagrams}, right diagram).

Similar to the charm-induced case, we apply an analogous strategy to examine the $[g \to {{\rm \Omega}_{3c}}]$ fragmentation channel.
As a preliminary illustration, we show the $z$ dependence of the gluon initial-scale FF adopted in the {\tt OMG3Q1.0} set.
To evaluate uncertainties at the starting scale, we perform a simplified, diagonal NLO DGLAP evolution using only the $P_{gg}$ kernel, following the same methodology employed in Ref.~\cite{Celiberto:2024mab} for gluon-induced tetraquark fragmentation.

The right panel of Fig.~\ref{fig:FFs_initial-scale} displays the $z$-weighted gluon FF, evaluated at $\mu_{F,0} = 6 m_c$, and evolved within the range $5m_c < \mu_F < 7m_c$.
As expected, the function vanishes at $z \to 0$ and $z \to 1$, and exhibits a pronounced peak around $z \simeq 0.25$.
Compared to charm FF, the gluon curve is wider and slightly shifted to lower $z$, reflecting a more uniform sharing of energy among final-state partons in gluon-initiated splittings.

We note that charm FF dominates by several orders of magnitude over its gluon counterpart.
This suppression is not merely an artifact of the scalar diquark model, rather it reflects a structural feature of fragmentation into triply heavy baryons.
While final states with an even number of heavy quarks, such as quarkonia or fully heavy tetraquarks, can naturally emerge from gluon splittings, the exclusive production of three heavy quarks in the valence state---with no accompanying light partons---poses a significant challenge for gluon fragmentation.

The scalar diquark framework, which presumes an $\mathcal{O}(\alpha_s)$ suppression compared to the quark-induced channel, emphasizes this aspect without rigidly imposing it.
Instead, it offers a phenomenologically motivated picture wherein the hierarchical suppression of gluon-induced baryon production is not imposed, but rather emerges as a natural outcome of the model.

This interpretation is further supported by comparing our findings for the $\rm \Omega_{3c}$ with results obtained for other exotic hadrons. 
As a remarkable example, while the lowest Fock state of the $\rm \Omega_{3c}$ is a valence $|ccc\rangle$ configuration, the corresponding state for a fully heavy tetraquark $\TQc$ is $|cc\bar{c}\bar{c}\rangle$. 
This structural difference has profound implications for gluon-induced fragmentation.

As will be shown in the right panel of Fig.~\ref{fig:NLO_FFs}, discussed in the following subsection, the behavior is reversed in the case of $\TQc$. 
The gluon-induced FF dominates over the quark-induced one by a factor ranging from 2 to 4, depending on the scale.
This contrasting trend originates from the fact that the tetraquark final state can be more naturally produced in gluon-initiated parton cascades. 
In particular, a high-energy gluon can radiate further gluons, which then split into heavy-quark/antiquark pairs, effectively producing configurations like $cc\bar{c}\bar{c}$. 
These can recombine into a compact color-singlet state, making the gluon channel particularly efficient for producing fully heavy tetraquarks.

By contrast, the baryonic final state $|ccc\rangle$ requires the exclusive production of three heavy quarks with no accompanying antiquarks or light partons, a configuration that is highly disfavored in gluon fragmentation. 
This comparison confirms that the observed suppression in the $\rm \Omega_{3c}$ gluon FF is not merely due to model-dependent features, but rather reflects a deeper structural limitation imposed by the quantum numbers and parton content of the final state.

In summary, the suppression of the gluon FF for $\rm \Omega_{3c}$ is not due to the diagrammatic structure alone, but rather to the unfavorable quantum number requirements of the final state, in particular the need to isolate three valence $c$ quarks without accompanying antiquarks or light partons. 
This conclusion is consistent with the behavior observed in other heavy-flavored exotic systems, where gluon-induced fragmentation is more efficient precisely because the final-state structure is more compatible with gluon splitting.

\subsection{The {\tt OMG3Q1.0} functions from {\HFNRevo}}
\label{ssec:FFs_OMG3Q10}

\begin{figure*}[!t]
\centering

\includegraphics[scale=0.650,clip]{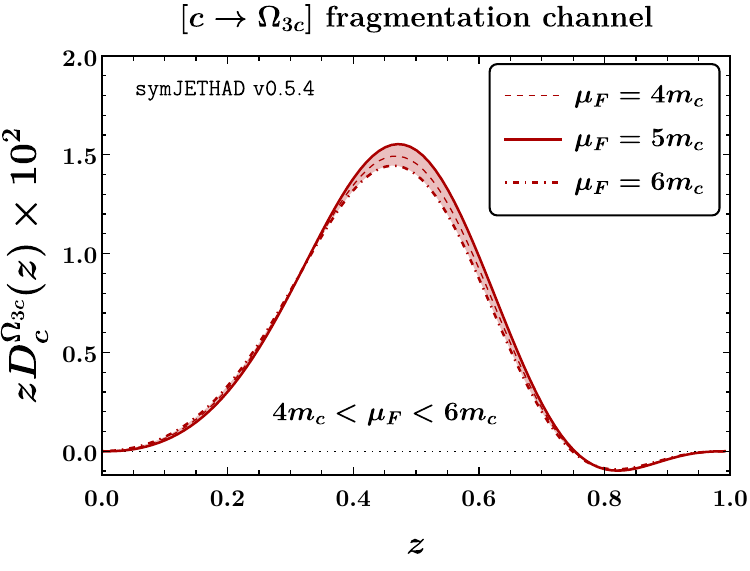}
\hspace{0.30cm}
\includegraphics[scale=0.650,clip]{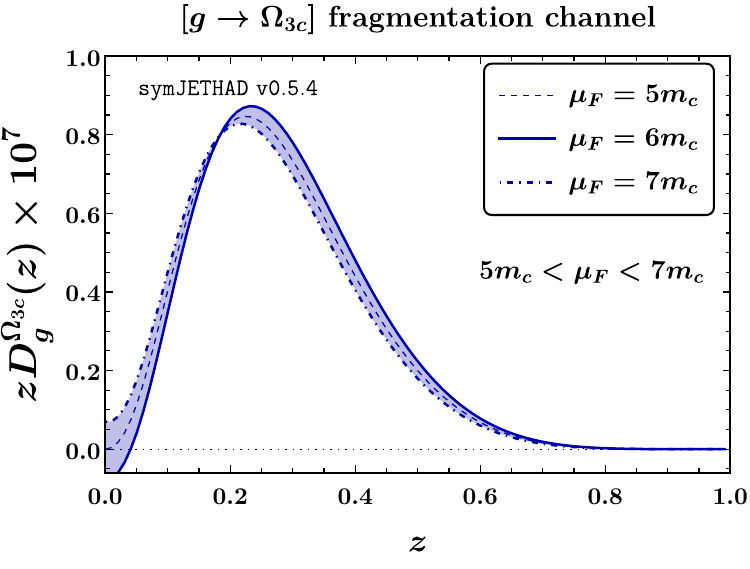}

\caption{Initial-scale charm (left) and gluon (right) channels to ${\rm \Omega}_{3c}$ for our {\tt OMG3Q1.0} FFs.
The shaded bands represent the effect of DGLAP evolution within the range $4 m_c < \mu_F < 6 m_c$ for the charm quark and $5 m_c < \mu_F < 7 m_c$ for the gluon.}
\label{fig:FFs_initial-scale}
\end{figure*}

To complete the construction of our {\tt OMG3Q1.0} collinear FFs for triply charmed ${\rm \Omega}$ baryons, we perform a proper DGLAP evolution of the initial-scale inputs presented before.
A key distinction from light-hadron fragmentation is that in this case both the (anti-)charm and gluon channels are subject to evolution thresholds.
This feature stems directly from the perturbative nature of partonic splittings into the lowest Fock state: gluon- and charm-initiated processes correspond to the right and left panels of Fig.~\ref{fig:O3Q_FF_diagrams}, respectively, and are governed by their associated SDCs.

As outlined earlier, kinematic constraints dictate that the minimum invariant mass required for quark fragmentation is $5m_c$, which we adopt as the threshold scale for the charm channel.
In contrast, gluon-induced production requires a higher threshold $6 m_c$, which we define as the starting scale for the (anti)charm FF.
To handle the evolution while properly incorporating these thresholds, we adopt a dedicated strategy based on the recently developed {\HFNRevo} scheme~\cite{Celiberto:2024mex,Celiberto:2024bxu,Celiberto:2024rxa}.

This framework is tailored to describe DGLAP evolution of heavy-hadron FFs starting from nonrelativistic inputs and relies on three core components: interpretation, evolution, and uncertainty quantification.
The first component allows us to interpret the low-transverse-momentum production mechanism as a two-parton fragmentation process, as outlined in Section~\ref{ssec:FFs_highlights}.
This forms the basis for a consistent matching between FFNS and VFNS calculations.
The third component provides a systematic approach to estimate the impact of missing higher-order uncertainties (MHOUs) by varying the evolution thresholds.

Although originally developed to bridge precision QCD predictions with a hadron-structure-oriented vision for quarkonium fragmentation, the {\HFNRevo} me\-thod\-ology has recently been extended to address exotic matter production, with promising results.
Applications to scalar ($0^{++}$) and tensor ($2^{++}$) $\TQc$ states~\cite{Celiberto:2024mab,Celiberto:2024beg}, as well as to rare baryons such as ${\rm \Omega}_{3c}$ (as explored in this work), have established {\HFNRevo} as a flexible framework for evolving FFs that involve both constituent heavy-quark and gluon initial-scale inputs.
This dual-channel structure introduces the need for a dedicated treatment of evolution thresholds, tailored to each partonic species.

In this exploratory study on ${\rm \Omega}_{3c}$ fragmentation, we postpone the implementation of matching procedures and uncertainty quantification.
Instead, we concentrate on developing a robust strategy for handling DGLAP evolution in the presence of quark and gluon thresholds.

According to {\HFNRevo}, the DGLAP evolution of heavy-hadron FFs is implemented in two successive stages.
In the context of rare baryon production, we begin by evolving the channel with the lowest threshold, namely the charm-initiated one.
The initial condition is provided by the (anti)charm FF at $\mu_{F,0} = 5 m_c$, as obtained in Section~\ref{ssec:FFs_c}.
This FF is then evolved up to $6 m_c$, the gluon threshold, using only the $P_{qq}$ kernel.
Since this evolution is both \emph{expanded} in powers of $\alpha_s$ and \emph{decoupled} from other partonic contributions, it can be performed analytically using the {\symJethad} plug-in~\cite{Celiberto:2020wpk,Celiberto:2022rfj,Celiberto:2023fzz,Celiberto:2024mrq,Celiberto:2024swu}.

Starting from this common scale, we apply a full numerical DGLAP evolution to produce the NLO {\tt OMG3Q1.0} set, which we release in {\tt LHAPDF} format.
We define the starting point of this step as the \emph{evolution-ready} scale, $Q_0$, corresponding to the highest threshold among active partonic species, $6 m_c$ in this case.
The $Q_0$ scale is the energy at which the numerical evolution is initialized.
For this purpose, we use {\tt APFEL++}~\cite{Bertone:2013vaa,Carrazza:2014gfa,Bertone:2017gds}.
In future developments, we also intend to interface with {\tt EKO}~\cite{Candido:2022tld,Hekhorn:2023gul} to further expand our evolution technology.

One could argue that our treatment is incomplete because of the omission of light- and bottom-quark channels.
To the best of our knowledge, no calculation currently exists for the collinear fragmentation of a nonconstituent quark into a triply heavy baryon.
As a result, in our two-step evolution framework, light and bottom quarks are not assigned any initial-scale FFs and instead emerge dynamically through DGLAP evolution.
However, drawing from analogies with NRQCD analyses of the color-singlet pseudoscalar~\cite{Braaten:1993rw,Braaten:1993mp,Artoisenet:2014lpa,Zhang:2018mlo,Zheng:2021mqr,Zheng:2021ylc} and vector~\cite{Braaten:1993rw,Braaten:1993mp,Zheng:2019dfk} charmonia, these channels are expected to be strongly suppressed compared to those initiated by gluons and charm quarks.

\begin{figure*}[!t]
\centering

   \hspace{-0.00cm}
   \includegraphics[scale=0.415,clip]{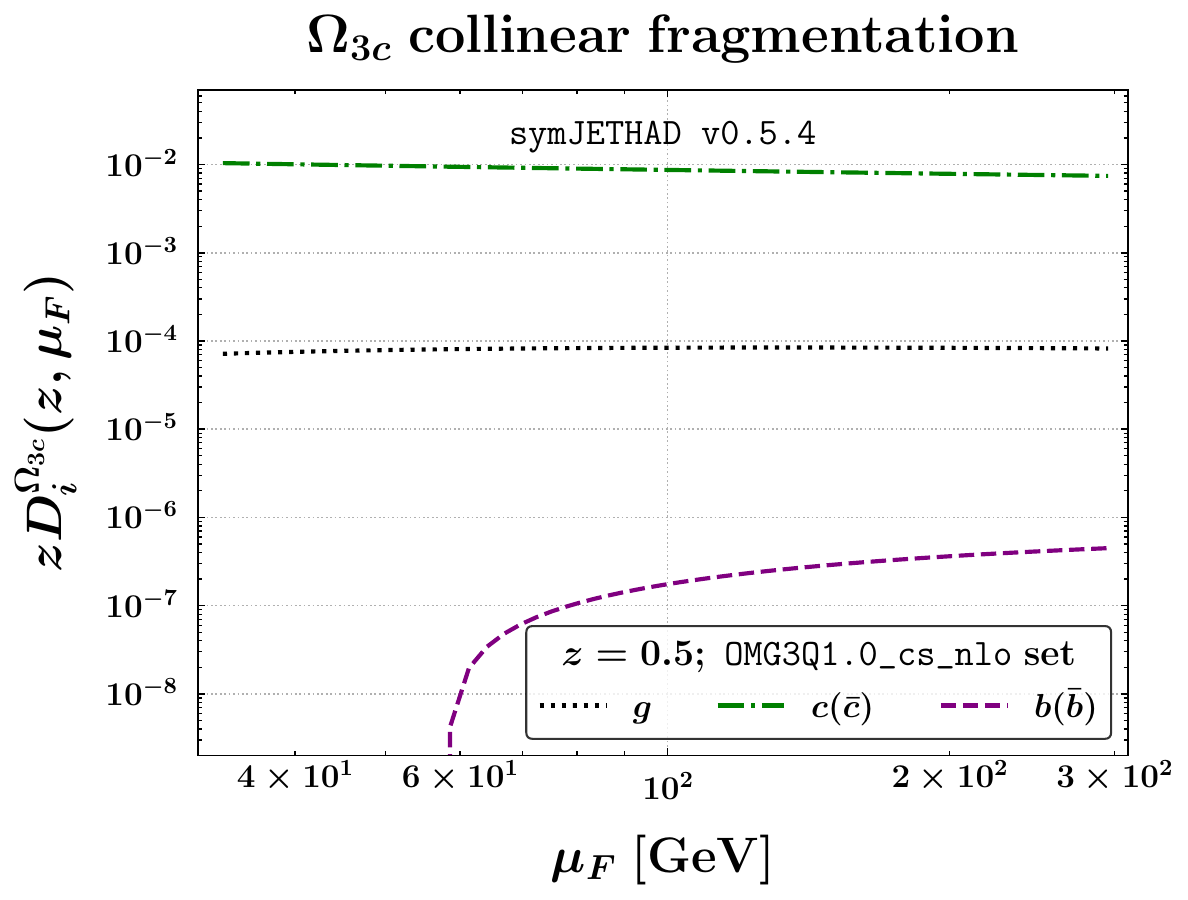}
   \hspace{-0.15cm}
   \includegraphics[scale=0.415,clip]{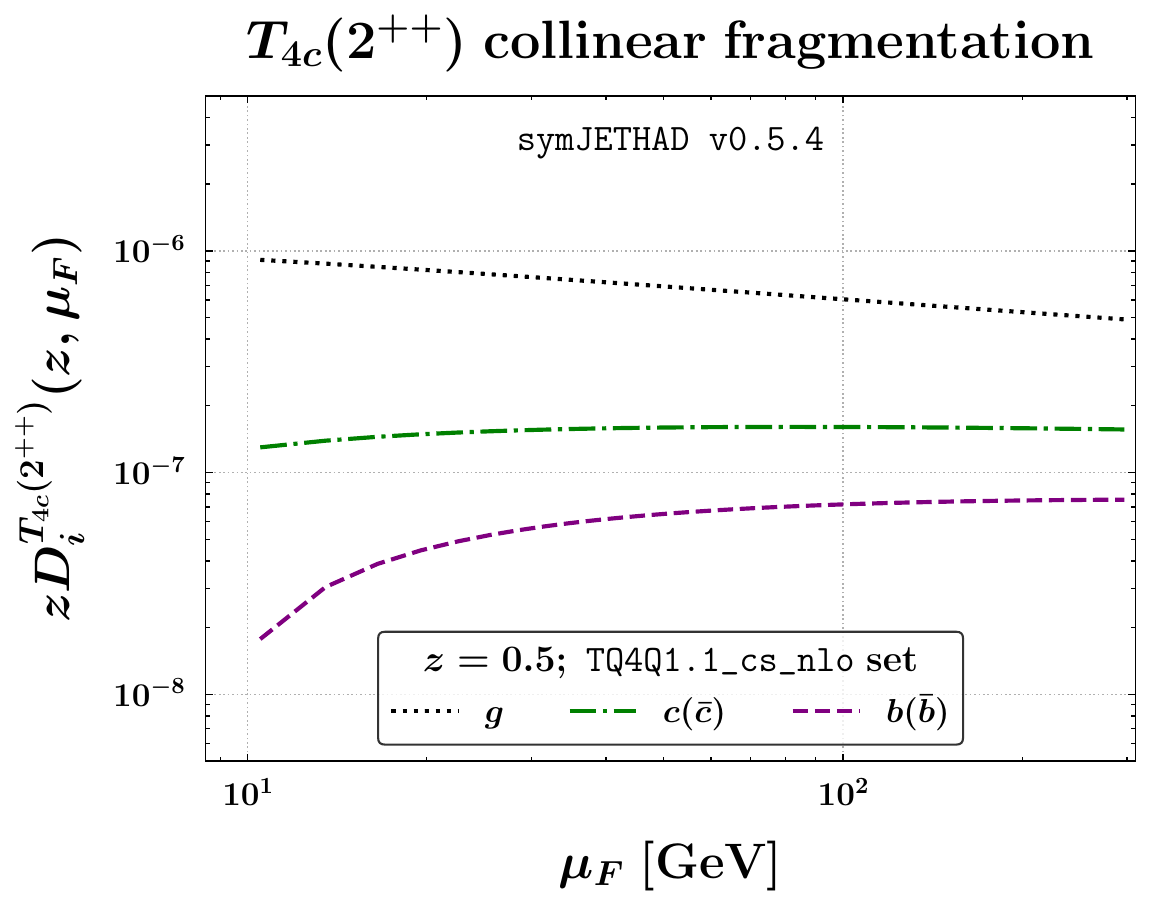}

\caption{Factorization-scale dependence of the {\tt OMG3Q1.0} FFs describing the VFNS fragmentation of the ${\rm \Omega}_{3c}$ baryon (left), compared with the corresponding {\tt TQ4Q1.1} FFs~\cite{Celiberto:2024beg,Celiberto:2025dfe,Celiberto:2025ziy} portraying the $\TQcTpp$ tetraquark (right), evaluated at $z = 0.5 \simeq \langle z \rangle$.}
\label{fig:NLO_FFs}
\end{figure*}

The plot on the left of Fig.~\ref{fig:NLO_FFs} displays the factorization scale dependence of the {\tt OMG3Q1.0} NLO functions that describe collinear VFNS fragmentation into ${\rm \Omega}_{3c}$ baryons.
For comparison, the panel on the right displays the corresponding $\mu_F$ evolution of the {\tt TQ4Q1.1} NLO functions~\cite{Celiberto:2024beg}, which describe the collinear VFNS fragmentation into the tensor $\TQc$ resonance ($2^{++}$), widely regarded as the leading candidate for the $X(6900)$ tetraquark~\cite{LHCb:2020bwg}.
For conciseness, both plots are evaluated at a representative value of the momentum fraction, $z = 0.5 \simeq \langle z \rangle$, which approximates the typical region where FFs provide dominant contributions to semihard final states~\cite{Celiberto:2016hae,Celiberto:2017ptm,Bolognino:2018oth,Celiberto:2020wpk,Celiberto:2021dzy,Celiberto:2021fdp,Celiberto:2022dyf,Celiberto:2022keu,Celiberto:2022kxx}.

A direct comparison of the results in Fig.~\ref{fig:NLO_FFs} reveals that the (anti)charm-to-${\rm \Omega}_{3c}$ fragmentation channel emerges as the dominant contribution.
It significantly outweighs the light-parton (not shown due to their negligible size) and bottom-quark channels across the entire $\mu_F$ range considered.
Moreover, it exceeds the gluon contribution by approximately 2 orders of magnitude.
This behavior contrasts with the case of the $\TQc(2^{++})$ state, where the gluon FF dominates over the charm FF, with ratios ranging from a factor of 5 up to an order of magnitude.
The prevailing role of the charm-initiated FF in ${\rm \Omega}_{3c}$ production directly reflects the initial-scale hierarchy between the two channels, as discussed earlier (see Fig.~\ref{fig:FFs_initial-scale}).
While the gluon FF increases with $\mu_F$ due to its feeding from timelike splittings, the charm FF exhibits a smoother trend and receives no comparable enhancement.
This explains why the initial five-order hierarchy between the charm and gluon FFs is reduced to roughly 2 orders of magnitude after DGLAP evolution.

Despite the charm FF being substantially larger in magnitude than its gluon counterpart, the latter still plays a key role in predicting ${\rm \Omega}_{3c}$ (semi-)inclusive production rates at hadron colliders.
This relevance stems from the overwhelming dominance of the gluon parton distribution function (PDF) over quark PDFs, which enhances the contribution of both the $[gg \to gg]$ partonic subprocess and the gluon-initiated fragmentation.
As a result, the simultaneous inclusion of initial-scale inputs for both the charm and gluon channels in our {\tt OMG3Q1.0} determinations strengthens the robustness and reliability of the proposed methodology.

We remark that the bottom-quark FF is not provided as an input at $Q_0$, but is dynamically generated via DGLAP evolution from gluon and charm channels. 
As a result, its shape and growth pattern with $\mu_F$ are strongly affected by the choice of the initial scale $Q_0$. 
In particular, the higher $Q_0$ value used in the $\rm \Omega_{3c}$ case delays the onset of the bottom component compared to the $T_{4c}$ case, where the evolution starts earlier. 
This effect accounts for the apparent difference in the slope of the $b$-FF curves shown in Fig.~\ref{fig:NLO_FFs}.

Finally, we observe that the gluon-to-${\rm \Omega}_{3c}$ FF exhibits a mild increase with $\mu_F$, while its gluon-to-$T{4c}$ counterpart displays a slow decrease.
Both behaviors are smoothly varying with respect to the factorization scale, a property that carries significant phenomenological implications.
Indeed, it has recently been shown that gluon FFs characterized by a smooth $\mu_F$ dependence act as effective ``stabilizers'' in high-energy resummed observables sensitive to the semi-inclusive production of singly~\cite{Celiberto:2021dzy,Celiberto:2021fdp} and multiply~\cite{Celiberto:2022dyf,Celiberto:2022keu,Celiberto:2023rzw} heavy-flavored hadrons.

This remarkable feature has been referred to as the \emph{natural stability}~\cite{Celiberto:2022grc} of high-energy resummation (see Section~\ref{sec:hybrid_factorization}).
The natural stability arising from the heavy-flavor fragmentation will constitute the cornerstone of our forthcoming phenomenological analysis (see Section~\ref{sec:results}).

\section{High-energy resummation at work}
\label{sec:hybrid_factorization}

The first part of this section (\ref{ssec:HE_resummation}) provides a concise overview of recent phenomenological advances in the exploration of the semihard regime of QCD. 
The second part (\ref{ssec:NLL_cross_section}) details the construction of the $\NLLp$ hybrid factorization framework and its adaptation to the semi-inclusive ${\rm \Omega}_{3c}$ plus jet hadroproduction process.

\subsection{The semihard regime of QCD}
\label{ssec:HE_resummation}

The production of heavy-flavored hadrons offers a valuable perspective on high-energy QCD, where energy logarithms become large enough to challenge perturbative expansions. The Balitsky-Fadin-Kuraev-Lipatov (BFKL) approach~\cite{Fadin:1975cb,Kuraev:1977fs,Balitsky:1978ic} addresses this by systematically resumming energy logarithms to all orders, covering both leading logarithmic (LL) terms $\alpha_s^n \ln (s)^n$ and next-to-leading logarithmic (NLL) corrections $\alpha_s^{n+1} \ln (s)^n$.

In the BFKL framework, cross sections are computed as transverse-momentum convolutions involving a universal NLO Green's function~\cite{Fadin:1998py,Ciafaloni:1998gs} and process-dependent, singly off-shell emission functions, also known as forward impact factors.
These functions incorporate collinear elements such as PDFs and FFs, leading to a \emph{hybrid-factorization} formalism that merges high-energy and collinear QCD dynamics.

Over the years, the BFKL resummation has been applied to a variety of processes, including Mueller-Navelet jets \cite{Mueller:1986ey,Ducloue:2013hia,Colferai:2015zfa,Celiberto:2015yba,Celiberto:2015mpa,Celiberto:2016ygs,Celiberto:2017ius,Caporale:2018qnm,deLeon:2021ecb,Celiberto:2022gji,Baldenegro:2024ndr}, dihadron~\cite{Celiberto:2016hae,Celiberto:2017ptm,Celiberto:2017ius,Celiberto:2020rxb,Celiberto:2022rfj} and hadron-jet systems~\cite{Bolognino:2018oth,Bolognino:2019cac,Bolognino:2019yqj,Celiberto:2020wpk,Celiberto:2020rxb,Mohammed:2022gbk,Celiberto:2022kxx}, multijets~\cite{Caporale:2016soq,Caporale:2016xku,Celiberto:2016vhn,Caporale:2016zkc,Celiberto:2017ius}, forward-Higgs~\cite{Hentschinski:2020tbi,Celiberto:2022fgx,Celiberto:2020tmb,Mohammed:2022gbk,Celiberto:2023rtu,Celiberto:2023uuk,Celiberto:2023eba,Celiberto:2023nym,Celiberto:2023rqp,Celiberto:2022zdg,Celiberto:2024bbv}, Drell-Yan~\cite{Celiberto:2018muu,Golec-Biernat:2018kem}, and heavy-flavor\-ed emissions~\cite{Celiberto:2017nyx,Boussarie:2017oae,Bolognino:2019ouc,Bolognino:2019yls,Celiberto:2021dzy,Celiberto:2021fdp,Celiberto:2022dyf,Celiberto:2023fzz,Celiberto:2022grc,Bolognino:2022paj,Celiberto:2022keu,Celiberto:2022kza,Celiberto:2024omj}. 
Single forward production rates have provided insights into small-$x$ gluon dynamics via the unintegrated gluon distribution (UGD), studied at HERA~\cite{Anikin:2011sa,Besse:2013muy,Bolognino:2018rhb,Bolognino:2018mlw,Bolognino:2019bko,Bolognino:2019pba,Celiberto:2019slj,Bolognino:2021bjd,Luszczak:2022fkf}, the Electron-Ion Collider (EIC)~\cite{Bolognino:2021niq,Bolognino:2021gjm,Bolognino:2021bjd,Bolognino:2022uty,Bolognino:2022ndh}.
This has subsequently enabled the development of resummed PDFs~\cite{Ball:2017otu,Abdolmaleki:2018jln,Bonvini:2019wxf,Silvetti:2022hyc,Silvetti:2023suu,Rinaudo:2024hdb} and improved low-$x$ TMD PDFs~\cite{Bacchetta:2020vty,Bacchetta:2024fci,Celiberto:2021zww,Bacchetta:2021oht,Bacchetta:2021lvw,Bacchetta:2021twk,Bacchetta:2022esb,Bacchetta:2022crh,Bacchetta:2022nyv,Celiberto:2022omz,Bacchetta:2023zir}.

In the context of heavy-hadron emissions, processes like ${\rm \Lambda}_c$~\cite{Celiberto:2021dzy} and $b$-hadron production~\cite{Celiberto:2021fdp} have revealed mechanisms to address challenges in natural-scale descriptions of semihard processes. Unlike light-particle emissions, which suffer from large NLL corrections and nonresummed threshold effects~\cite{Ducloue:2013bva,Caporale:2014gpa,Bolognino:2018oth,Celiberto:2020wpk}, heavy-flavored hadrons exhibit a \emph{natural stabilization} trend~\cite{Celiberto:2022grc}, driven by collinear VFNS fragmentation.

These findings have inspired further studies, including the development of VFNS DGLAP-evolving FFs based on NRQCD inputs~\cite{Braaten:1993mp,Zheng:2019dfk,Braaten:1993rw,Chang:1992bb,Braaten:1993jn,Ma:1994zt,Zheng:2019gnb,Zheng:2021sdo,Feng:2021qjm,Feng:2018ulg}, extending from vector quarkonia~\cite{Celiberto:2022dyf,Celiberto:2023fzz} to $\BCs$ and $\Bss$ mesons \cite{Celiberto:2022keu,Celiberto:2024omj}. 
Natural stability has also opened a portal to exotic matter, serving as a phenomenological playground to study the fragmentation of the leading power doubly~\cite{Celiberto:2023rzw,Celiberto:2024beg} or fully heavy tetraquarks~\cite{Celiberto:2024mab,Celiberto:2024beg,Celiberto:2025dfe,Celiberto:2025ziy} and pentacharms~\cite{Celiberto:2025ipt}.

\begin{figure*}[!t]
\centering

\includegraphics[width=0.575\textwidth]{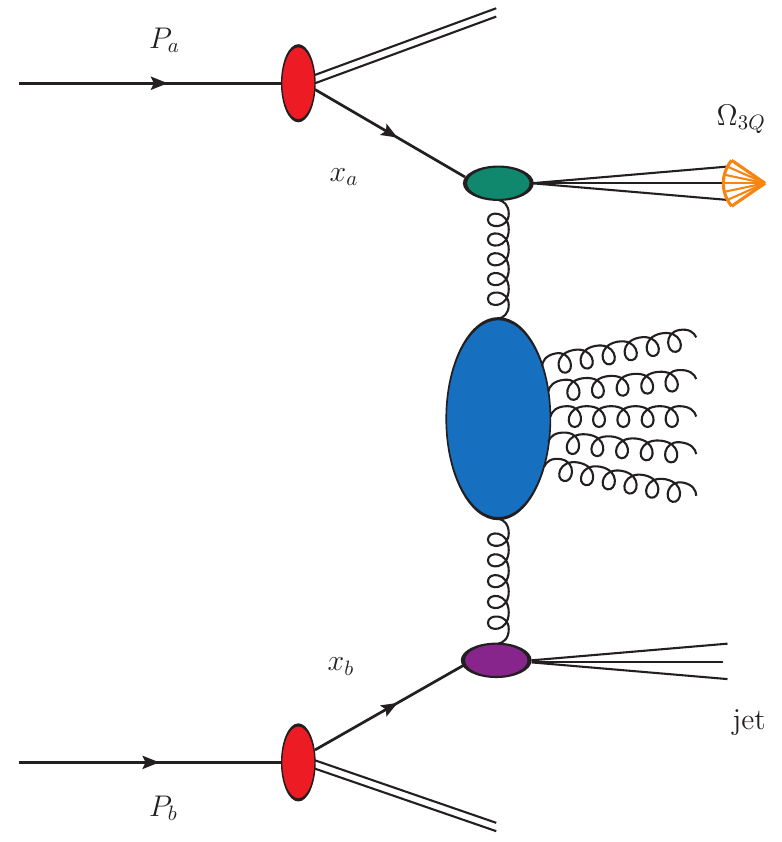}

\caption{Pictorial representation of semi-inclusive hadroproduction of a ${\rm \Omega}_{3Q}$ baryon plus a jet within the hybrid factorization. Red blobs depict collinear PDFs. The singly off-shell coefficient function embodied in the hadron (jet) emission function is portrayed by the green (violet) oval.
The orange composite arrow represents the inclusive formation of a ${\rm \Omega}_{3Q}$ baryon. The large blue oval blob stands for the high-energy Green's function resumming secondary gluon emissions in the $t$-channel. Diagram was realized with the {\tt JaxoDraw 2.0} code~\cite{Binosi:2008ig}.}
\label{fig:pictorial}
\end{figure*}

\subsection{Differential cross section at NLL/NLO$^+$}
\label{ssec:NLL_cross_section}

As an application to hadron-collider phenomenology, we examine the following semi-inclusive reaction (see Fig.~\ref{fig:pictorial})
\begin{equation}
\label{eq:process}
 {\rm p}(P_a) \, {\rm p}(P_b) \to {{\rm \Omega}_{3c}}(q_1, y_1, \phi_1) + {\cal X} + {\rm jet}(q_2, y_2, \phi_2) \;,
\end{equation}
where a triply charmed baryon ${\rm \Omega}_{3c}$ is detected with four-momentum $q_1$, rapidity $y_1$, and azimuthal angle $\phi_1$. In addition, a light jet is tagged with four-momentum $q_2$, rapidity $y_2$, and azimuthal angle $\phi_2$. Both objects feature high transverse momenta, such that $|\vec q_{1,2}| \gg {\rm \Lambda}_{\rm QCD}$, and a large rapidity distance, $\DY \equiv y_1 - y_2$. An undetected final-state gluon system, ${\cal X}$, is produced inclusively. We express the final-state transverse momenta in the Sudakov-vector basis, defined by the parent protons' momenta, $P_{a,b}$, where $P_{a,b}^2 = 0$ and $({P_a} \cdot {P_b}) = s/2$. 
This leads to the decomposition
\begin{equation}
\label{Sudakov}
 q_{1,2} = x_{1,2} \, P_{a,b} - \frac{q_{1,2 \perp}^2}{x_{1,2} s} \, P_{b,a} + q_{{1,2 \perp}} \;,
\end{equation}
with $q_{1,2 \perp}^2 \equiv - \vec q_{1,2}^{\,2}$.
In the center-of-mass frame, the following relations hold between the final-state longitudinal momentum fractions, rapidities, and transverse momenta:
\begin{equation}
\label{xyp}
 x_{1,2} = \frac{|\vec q_{1,2}|}{\sqrt{s}} e^{\pm y_{1,2}} \;, \qquad \drv y_{1,2} = \pm \frac{d x_{1,2}}{x_{1,2}} \;,
\end{equation}
and therefore
\begin{equation}
\label{Delta_Y}
 \DY \equiv y_1 - y_2 = \ln\frac{x_1 x_2 s}{|\vec q_1| |\vec q_2|} \;.
\end{equation}

In a purely collinear factorization approach, the LO differential cross section for these reactions can be written as a one-dimensional convolution involving the on-shell hard factor, proton PDFs, and ${\rm \Omega}_{3c}$ FFs:
\begin{eqnarray}
\label{sigma_collinear}
&&\hspace{-0.25cm}
\frac{\drv\sigma^{\rm LO}_{\rm [collinear]}}{\drv x_1\drv x_2\drv ^2\vec q_1\drv ^2\vec q_2}
= \hspace{-0.25cm} \sum_{m,n=q,{\bar q},g}\int_0^1 \hspace{-0.20cm} \drv x_a \!\! \int_0^1 \hspace{-0.20cm} \drv x_b\ f_m\left(x_a\right) f_n\left(x_b\right) 
\nonumber \\
&&\quad\times \, 
\int_{x_1}^1 \hspace{-0.15cm} \frac{\drv \zeta}{\zeta} \, D^{{\rm \Omega}_{3c}}_m\left(\frac{x_1}{\zeta}\right) 
\frac{\drv {\hat\sigma}_{m,n}\left(\hat s\right)}
{\drv x_1\drv x_2\drv ^2\vec q_1\drv ^2\vec q_2}\;,
\end{eqnarray}
where the indices $m,n$ sum over all partons except the (anti)top quark, which does not participate in hadronization. 
For simplicity, the explicit dependence on the factorization scale $\mu_F$ is omitted in Eq.~\eqref{sigma_collinear}. 
The functions $f_{m,n}\left(x_{a,b}, \mu_F \right)$ represent the proton collinear PDFs, while the $D^{{\rm \Omega}_{3c}}_m\left(x_1/\zeta, \mu_F \right)$ functions are the ${\rm \Omega}_{3c}$ collinear FFs. 
The variables $x_{a,b}$ correspond to the longitudinal-momentum fractions of the parent partons, and $\zeta$ is the momentum fraction of the outgoing parton fragmenting into the ${\rm \Omega}_{3c}$ particle. 
Finally, $\drv\hat\sigma_{m,n}\left(\hat s\right)$ is the partonic hard factor, where $\hat s \equiv x_a x_b s$ is the partonic center-of-mass energy squared.

Conversely, the differential cross section within our hybrid high-energy and collinear factorization approach is expressed as a transverse-momentum convolution involving the BFKL Green's function and the two singly off-shell emission functions. 
We rewrite the cross section as a Fourier series of the azimuthal-angle coefficients, ${\cal C}_{k \, \ge \, 0}$.
We write
\begin{equation}
 \label{dsigma_Fourier}
 \hspace{-0.22cm}
 \frac{(2\pi)^2 \, \drv \sigma}{\drv y_1 \drv y_2 \drv |\vec q_1| \drv |\vec q_2| \drv \phi_1 \drv \phi_2} \!=\!
 \left[ {\cal C}_0 + 2 \! \sum_{k=1}^\infty \! \cos (k \phi)\,
 {\cal C}_k \right]\, ,
\end{equation}
where $\phi \equiv \phi_1 - \phi_2 - \pi$.

In the $\MSb$ renormalization scheme and using the BFKL resummation, we obtain (see Ref.~\cite{Caporale:2012ih} for more details)
\begin{equation}
\label{Ck_NLLp_MSb}
 \CkNLLp = 
 \frac{e^{\DY}}{s}
 \int_{-\infty}^{+\infty} \!\!\! \drv \nu \, e^{{\DY} \bar \alpha_s(\mu_R)
 \chi^{\rm NLO}(k,\nu)} 
\end{equation}
\[
 \times \, \alpha_s^2(\mu_R) \biggl\{ \E_{{\rm \Omega}_{3c}}^{\rm NLO}(k,\nu,|\vec q_1|, x_1)[\E_J^{\rm NLO}(k,\nu,|\vec q_2|,x_2)]^* 
\]
\[
 +
 \left.
 \alpha_s^2(\mu_R) \DY \frac{\beta_0}{4 \pi} \,
 \chi(k,\nu)
 \left[\ln\left(|\vec q_1| |\vec q_2|\right) + \frac{i}{2} \, \frac{\drv}{\drv \nu} \ln\frac{\E_{{\rm \Omega}_{3c}}}{\E_J^*}\right]
 \right\}
 .
\]
In this equation, $\bar \alpha_s(\mu_R) \equiv \alpha_s(\mu_R) N_c/\pi$ is the QCD running coupling, with $N_c = 3$ representing the number of colors, and $\beta_0 = 11N_c/3 - 2 n_f/3$ being the first coefficient of the QCD $\beta$ function. 
We choose a two-loop running coupling setup with $\alpha_s\left(m_Z\right)=0.118$ and the dynamic number of flavors $n_f$.

The high-energy kernel in the exponent of Eq.~\eqref{Ck_NLLp_MSb} captures the resummation of energy logarithms at NLL accuracy:
\begin{eqnarray}
 \label{chi}
 \chi^{\rm NLO}(k,\nu) = \chi(k,\nu) + \bar\alpha_s \hat \chi(k,\nu) \;,
\end{eqnarray}
where $\chi(k,\nu)$ is the eigenvalue of the LO kernel:
\begin{eqnarray}
 \label{kernel_LO}
 \chi\left(k,\nu\right) = -2\gamma_{\rm E} - 2 \, {\rm Re} \left\{ \psi\left(\frac{1+k}{2} + i \nu \right) \right\} \, ,
\end{eqnarray}
with $\gamma_{\rm E}$ being the Euler-Mascheroni constant and $\psi(z) \equiv \Gamma^\prime(z)/\Gamma(z)$ the logarithmic derivative of the $\Gamma$ function. The function $\hat\chi(k,\nu)$ in Eq.~\eqref{chi} represents the NLO correction to the high-energy kernel,
\begin{eqnarray}
\label{chi_NLO}
&\hat \chi&\left(k,\nu\right) = \bar\chi(k,\nu)+\frac{\beta_0}{8 N_c}\chi(k,\nu)
\\ \nonumber &\times& \left\{-\chi(k,\nu)+10/3+2\ln\left[\left(\mu_R^2/(|\vec q_1| |\vec q_2|\right)\right]\right\} \;,
\end{eqnarray}
with the characteristic $\bar\chi(k,\nu)$ function calculated in Ref.~\cite{Kotikov:2000pm}.

The two quantities  
\begin{eqnarray}  
\label{EFs}  
\E_{{\rm \Omega}_{3c},J}^{\rm NLO}(k,\nu,|\vec q_{1,2}|,x_{1,2}) =  
\E_{{\rm \Omega}_{3c},J} +  
\alpha_s(\mu_R) \, \hat \E_{{\rm \Omega}_{3c},J}  
\end{eqnarray}  
represent the NLO emission functions for the rare baryon (${\rm \Omega}_{3c}$) and the light jet ($J$), respectively. 
These objects are calculated in Mellin space and projected onto the eigenfunctions of the LO kernel. 
For the ${\rm \Omega}_{3c}$ emission function, we adopt the NLO calculation from Ref.~\cite{Ivanov:2012iv}. 
Although originally tailored for light-flavored hadron production, this computation aligns with our VFNS approach for heavy baryons, provided that transverse momenta exceed the heavy-quark thresholds relevant for DGLAP evolution.  

At the LO level, the ${\rm \Omega}_{3c}$ emission function reads 
\begin{equation}  
\begin{split}  
 \label{LOOMGEF}  
 \E_{{\rm \Omega}_{3c}}(k,\nu,|\vec q_1|,x_1) = \delta_c \, |\vec q_1|^{2i\nu-1} \int_x^1 \frac{\drv \zeta}{\zeta} \; \hat{x}^{1-2i\nu} \\  
 \times \, \Big[\rho_c f_g(\zeta)D_g^{{\rm \Omega}_{3c}}\left(\hat{x}\right)  
 + \sum_{m=q,\bar q} f_m(\zeta)D_m^{{\rm \Omega}_{3c}}\left(\hat{x}\right)\Big] \;,  
\end{split}  
\end{equation}  
where $\hat{x} = x/\zeta$, $\delta_c = 2 \sqrt{C_F/C_A}$, and $\rho_c = C_A/C_F$. Here, $C_F = (N_c^2-1)/(2N_c)$ and $C_A = N_c$ are the Casimir factors associated with gluon emissions from quarks and gluons, respectively. The complete NLO formula for $\E_{{\rm \Omega}_{3c}}^{\rm NLO}$ is available in Ref.~\cite{Ivanov:2012iv}.  

The LO jet emission function is expressed as  
\begin{equation}  
 \label{LOJEF}  
 \!\!\E_J(k,\nu,|\vec q_2|,x) = \delta_c  
 |\vec q_2|^{2i\nu-1} \!\Big[\rho_c f_g(x)  
 + \!\!\!\sum_{n=q,\bar q} f_n(x)\Big] \,.  
\end{equation}  
The corresponding NLO correction can be obtained by combining Eq.~(36) of Ref.~\cite{Caporale:2012ih} with Eqs.~(4.19) and~(4.20) of Ref.~\cite{Colferai:2015zfa}. These calculations are grounded in results from Refs.~\cite{Ivanov:2012iv,Ivanov:2012ms}, optimized for numerical analyses. They employ a \emph{small-cone} jet selection function~\cite{Furman:1981kf,Aversa:1988vb} based on a \emph{cone-type} algorithm~\cite{Colferai:2015zfa}.  

Equations~(\ref{Ck_NLLp_MSb}),~(\ref{LOOMGEF}), and~(\ref{LOJEF}) clearly demonstrate the implementation of our hybrid high-energy and collinear factorization scheme.  
In this approach, the cross section is expressed through the BFKL formalism, with the Green's function and emission functions as key components. 
The Green's function handles the resummation of large logarithmic contributions in the high-energy limit, while the emission functions encapsulate the PDFs and FFs, effectively bridging collinear factorization with high-energy dynamics.  

The `$+$' superscript in the $\CkNLLp$ label signifies that the expression of azimuthal coefficients in Eq.~(\ref{Ck_NLLp_MSb}) incorporates corrections beyond the NLL accuracy. 
These enhancements stem from two key sources: the exponentiated NLO corrections to the high-energy kernel and the interplay between the NLO corrections to the impact factors. 
As a result, the azimuthal coefficients provide a more refined representation, capturing intricate effects essential for precise predictions in processes where both high-energy and collinear logarithms are influential. 

By discarding all NLO contributions in Eq.~\eqref{Ck_NLLp_MSb}, one retrieves the pure LL limit of the angular coefficients, thus having 
\begin{equation}
\begin{split}
 \label{Ck_LL_MSb}
 &\CkLL = 
 \frac{e^{\DY}}{s}
 \int_{-\infty}^{+\infty} \drv \nu \, e^{{\DY} \bar \alpha_s(\mu_R)\chi(k,\nu)}
 \\[0.18cm] &\hspace{0.15cm}
 \times \, \alpha_s^2(\mu_R) \, \E_{{\rm \Omega}_{3c}}(k,\nu,|\vec q_1|, x_1)[\E_J(k,\nu,|\vec q_2|,x_2)]^* \;.
\end{split}
\end{equation}  

Then, to get a meaningful comparison between high-energy resummed predictions and those derived from a purely collinear, DGLAP-inspired framework, it becomes crucial to evaluate observables within both our hybrid factorization scheme and fixed-order computations. 
Unfortunately, current limitations mean that no numerical tools exist to calculate fixed-order distributions at NLO for inclusive semihard hadron plus jet production. 
To address this gap and quantify the influence of high-energy resummation on DGLAP predictions, we adopt an alternative strategy.

Our methodology, originally developed to investigate the angular distributions of Mueller-Navelet~\cite{Celiberto:2015yba,Celiberto:2015mpa} and hadron-jet~\cite{Celiberto:2020wpk} azimuthal distributions, involves truncating the high-energy expansion at the NLO level. 
This approach allows us to simulate the high-energy behavior as it would appear in a purely NLO calculation. 
Specifically, we achieve this by limiting the expansion of the azimuthal coefficients to ${\cal O}(\alpha_s^3)$, effectively constructing a high-energy fixed-order ($\HENLOp$) approximation. 
This approximation provides a practical framework for our phenomenological studies, facilitating a systematic comparison of the BFKL resummation effects with the high-energy regime of fixed-order predictions.  
The $\HENLOp$ angular coefficients, computed within the $\MSb$ renormalization scheme, are given by  
\begin{align}
\label{Ck_HENLOp_MSb}
 &\CkHENLOp = 
 \frac{e^{\DY}}{s} 
 \int_{-\infty}^{+\infty} \drv \nu \, 
 \alpha_s^2(\mu_R)
 \nonumber \\[0.75em]
 &\hspace{0.50cm}\times \,
 \left[ 1 + \bar \alpha_s(\mu_R) \DY \chi(k,\nu) \right]
 \\[0.75em] \nonumber
 &\hspace{0.50cm}\times \,
 \E_{{\rm \Omega}_{3c}}^{\rm NLO}(k,\nu,|\vec q_1|, x_1)[\E_J^{\rm NLO}(k,\nu,|\vec q_2|,x_2)]^* \;.
\end{align}  

In our analysis, the factorization scale ($\mu_F$) and renormalization one ($\mu_R$) are set to \emph{natural} energy values determined by the kinematics of the final state. 
We define $\mu_F = \mu_R \equiv \mu_{\cal N}$, where the natural scale is $\mu_{\cal N} = m_{{\rm \Omega}_{3c}, \perp} + |\vec  q_1|$.
Here, $m_{{\rm \Omega}_{3c}, \perp} = \sqrt{m_{{\rm \Omega}_{3c}}^2 + |\vec q_1|^2}$ represents the transverse mass of the fully heavy baryon, and we set $m_{{\rm \Omega}_{3c}} = 3m_c$. 
The transverse mass of the light jet is equal to its transverse momentum, $|\vec{q}_2|$.

Although the emission of two particles naturally involves two distinct energy scales, we simplify the analysis by combining these into a single reference scale ($\mu_{\cal N}$), defined as the sum of the transverse masses of both particles. This choice aligns with strategies employed in other precision QCD calculations and codes, such as those in Refs.~\cite{Alioli:2010xd,Campbell:2012am,Hamilton:2012rf}. It enables a consistent comparison of our results with predictions from different approaches while adhering to common conventions in QCD phenomenology.

To investigate the impact of MHOUs, we vary both $\mu_F$ and $\mu_R$ within a range of $\mu_{\cal N}/2$ to $2\mu_{\cal N}$, controlled by the parameter $C_\mu$. 
This method allows us to evaluate the sensitivity of our predictions to scale variations, providing a robust estimation of theoretical uncertainties.

\section{From HL-LHC to FCC}
\label{sec:results}

All numerical results presented in this work were obtained using the \textsc{Python}+\textsc{Fortran} {\Jethad} multimodular interface~\cite{Celiberto:2020wpk,Celiberto:2022rfj,Celiberto:2023fzz,Celiberto:2024mrq}.
For the proton PDFs, we used the {\tt NNPDF4.0} NLO determination~\cite{NNPDF:2021uiq,NNPDF:2021njg}, accessed through {\tt LHAPDF v6.5.5}~\cite{Buckley:2014ana}.

The uncertainty bands in the plots reflect both the impact of MHOUs and errors arising from multidimensional numerical integrations, which were kept consistently below 1\% by the {\Jethad} integrators.

\subsection{Rapidity distributions}
\label{ssec:DY}

The first production rate matter of our phenomenological analysis is the rapidity distribution, namely the cross section differential in the rapidity interval, $\DY = y_1 - y_2$, between the baryon and the jet
\begin{equation}
\begin{split}
\label{DY_distribution}
 \hspace{-0.12cm}
 \frac{\drv \sigma(\DY, s)}{\drv \DY} &=
 \int_{y_1^{\rm min}}^{y_1^{\rm max}} \!\!\!\!\! \drv y_1
 \int_{y_2^{\rm min}}^{y_2^{\rm max}} \!\!\!\!\! \drv y_2
 \, \,
 \delta (y_1 - y_2 - \DY)
 \\[0.10cm]
 &\times \,
 \int_{|\vec q_1|^{\rm min}}^{|\vec q_1|^{\rm max}} 
 \!\!\!\!\! \drv |\vec q_1|
 \int_{|\vec q_2|^{\rm min}}^{|\vec q_2|^{\rm max}} 
 \!\!\!\!\! \drv |\vec q_2|
 \, \,
 {\cal C}_{0}^{\rm [accuracy]}
 \;,
\end{split}
\end{equation}
with ${\cal C}_{0}$ being the $k=0$, $\phi$-averaged angular coefficient given in Section~\ref{ssec:NLL_cross_section}.
In this context, the `${\rm [accuracy]}$' superscript includes $\NLLp$, $\LL$, or $\HENLOp$. 

The transverse momenta of the heavy baryon span from $60$ to $120$~GeV, while those of the jet range from $50$ to $120$~GeV. 
These ranges align with current and prospective analyses of jets and hadrons at the LHC~\cite{Khachatryan:2016udy,Khachatryan:2020mpd}.
Using \emph{asymmetric} windows for the observed transverse momenta enhances the prominence of the high-energy signal over the fixed-order background~\cite{Celiberto:2015yba,Celiberto:2015mpa,Celiberto:2020wpk}.

Our choice of rapidity intervals aligns with established criteria in ongoing LHC studies. Baryon detections, restricted to the barrel calorimeter as in the CMS experiment~\cite{Chatrchyan:2012xg}, are limited to the rapidity range $|y_1| < 2.5$. 
For jets, which can also be tagged in the end cap regions~\cite{Khachatryan:2016udy}, we consider a broader range of rapidity of $|y_2| < 4.7$.

\begin{figure*}[!t]
\centering

   \hspace{0.00cm}
   \includegraphics[scale=0.400,clip]{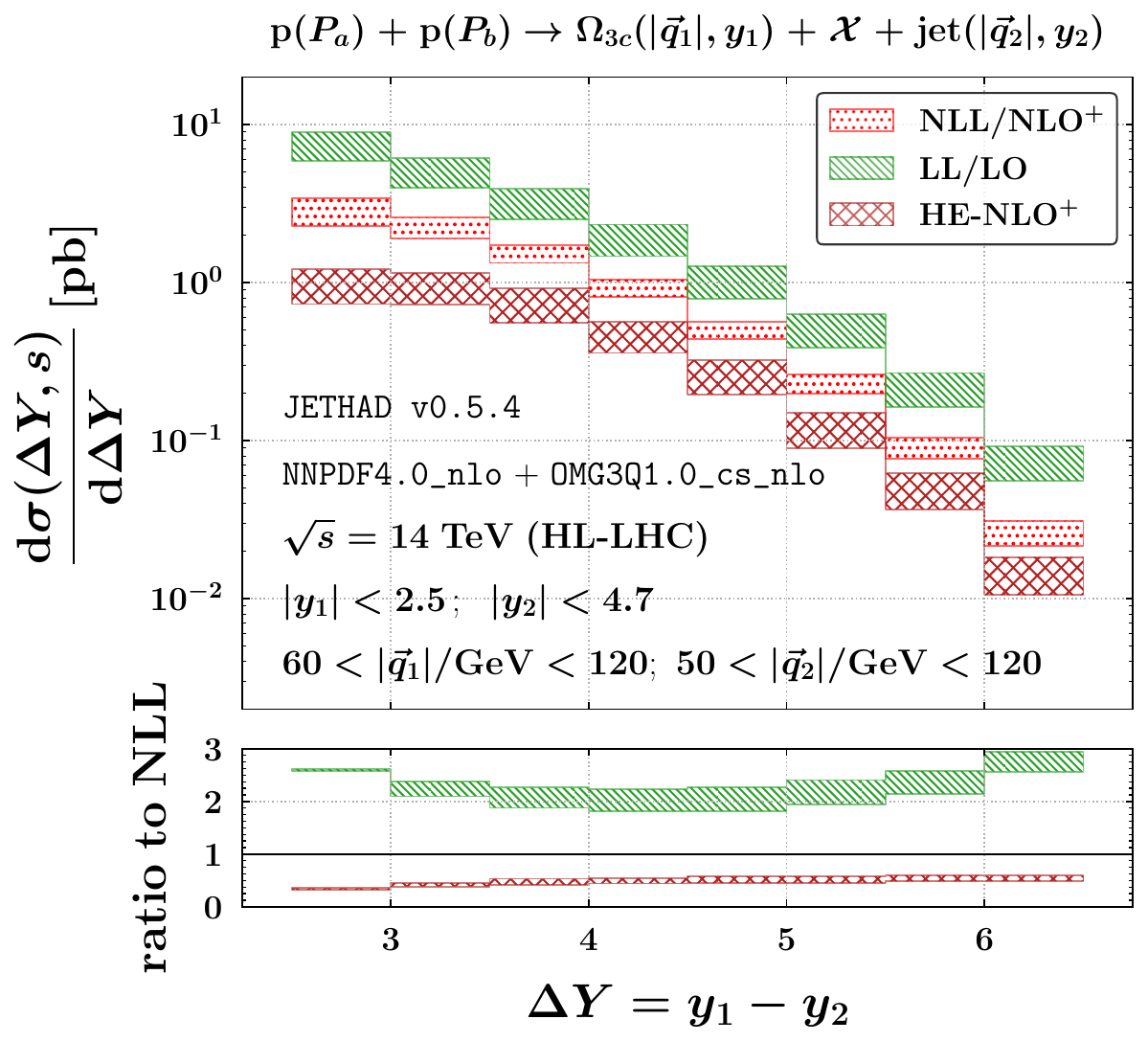}
   \hspace{0.20cm}
   \includegraphics[scale=0.400,clip]{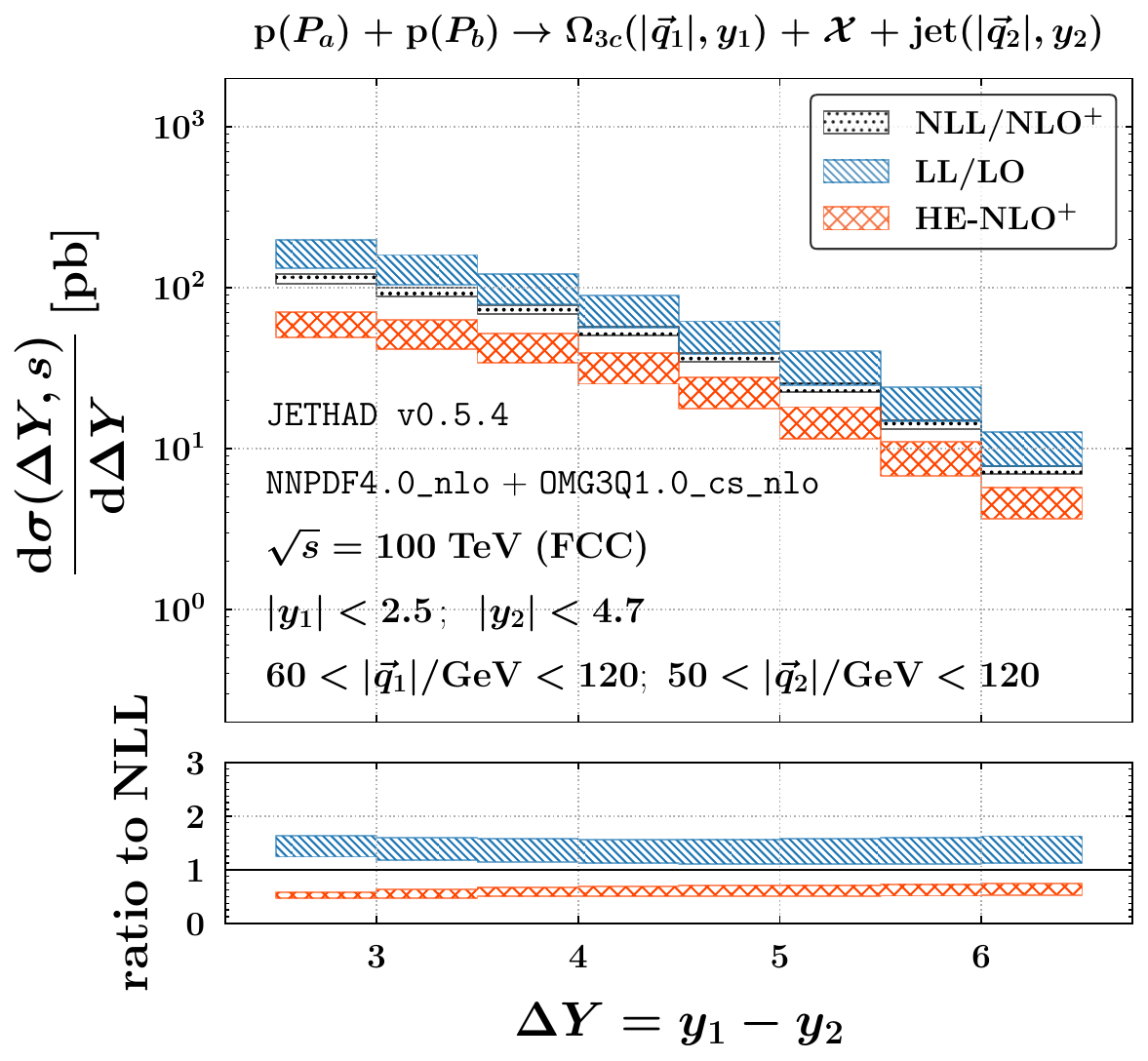}

\caption{$\DY$ differential distribution for semi-inclusive ${\rm \Omega}_{3c}$ plus jet detections at $\sqrt{s} = 14$~TeV (HL-LHC, left) and $100$~TeV (nominal FCC, right).
{\tt NNPDF4.0} NLO proton PDFs~\cite{NNPDF:2021uiq,NNPDF:2021njg} are used in combination with {\tt OMG3Q1.0} NLO heavy-baryon FFs~\cite{Celiberto:2025_OMG3Q10}.
Ancillary panels below the main plots show the ratio of $\LL$ and $\HENLOp$ predictions to $\NLLp$.
Uncertainty bands account for the combined effects of MHOUs and numerical phase-space integration.}
\label{fig:rapidity_O3c}
\end{figure*}

Figure~\ref{fig:rapidity_O3c} presents our resummed predictions for ${\rm \Omega}_{3c}$ plus jet $\DY$ distributions at $\sqrt{s} = 14$~TeV (HL-LHC, left panel) and $\sqrt{s} = 100$~TeV (FCC, right panel).
To facilitate meaningful comparisons with prospective experimental measurements, we adopt uniformly spaced $\DY$ bins of fixed width, $\DY = 0.5$.
Ancillary panels beneath the main plots show ratios of the $\LL$ and $\HENLOp$ predictions to the $\NLLp$ baseline.

The resulting cross sections are encouraging, spanning roughly from $10^{2}$~pb down to $10^{-2}$~pb.
Notably, a more than tenfold increase is observed when moving from HL-LHC to FCC energies.
We emphasize that the vertical scales of the two panels differ accordingly.
All $\DY$ spectra display a monotonic decreasing behavior with increasing $\DY$.
This trend reflects the interplay of two competing mechanisms.
The partonic hard factor grows with energy (and thus with $\DY$), in line with expectations from high-energy resummation.
Conversely, this growth is counteracted by suppression due to the collinear convolution with PDFs and FFs in the emission functions (see Eqs.~\eqref{LOOMGEF} and~\eqref{LOJEF}).

\begin{figure*}[!t]
\centering

   \hspace{0.00cm}
   \includegraphics[scale=0.400,clip]{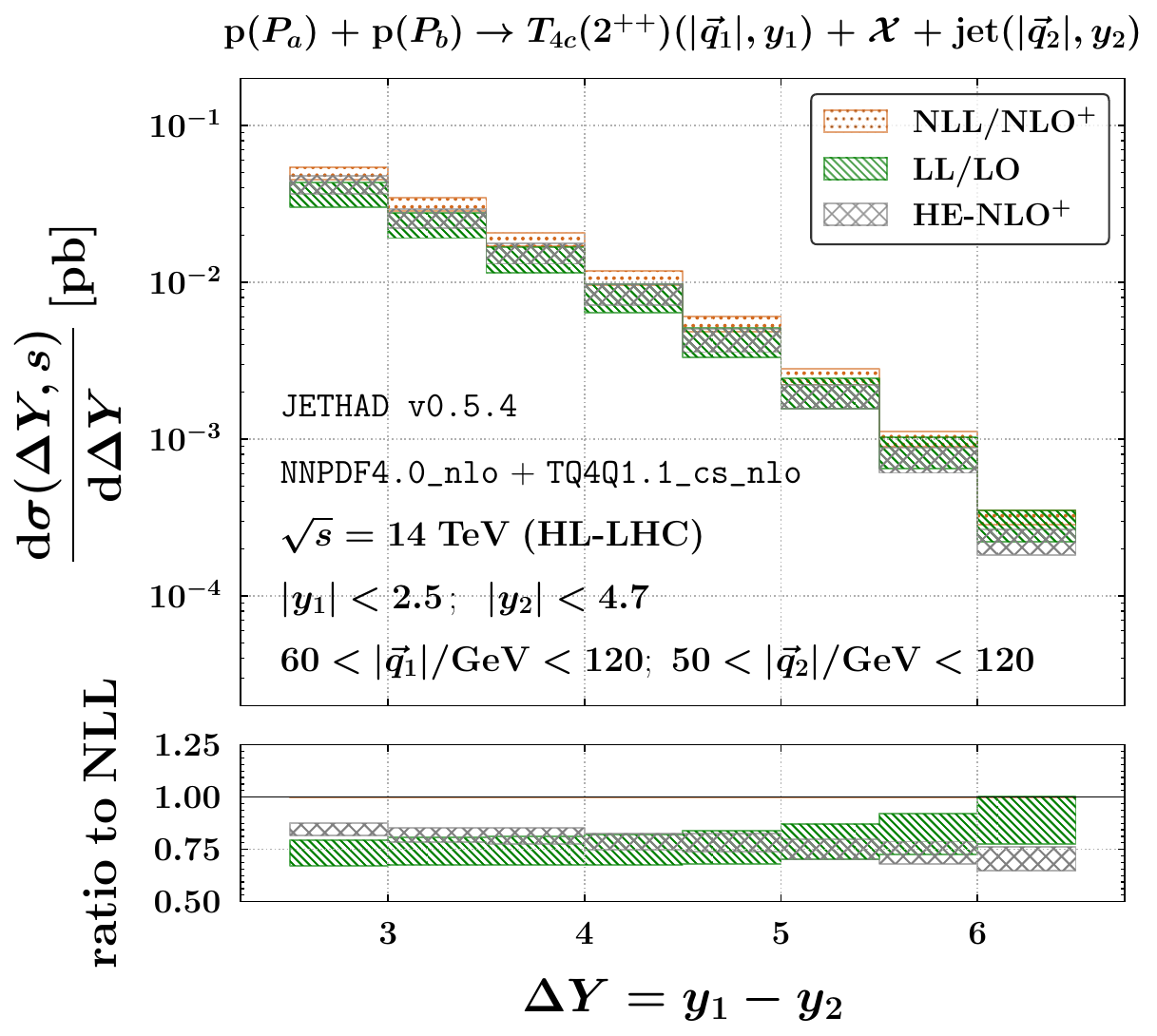}
   \hspace{0.20cm}
   \includegraphics[scale=0.400,clip]{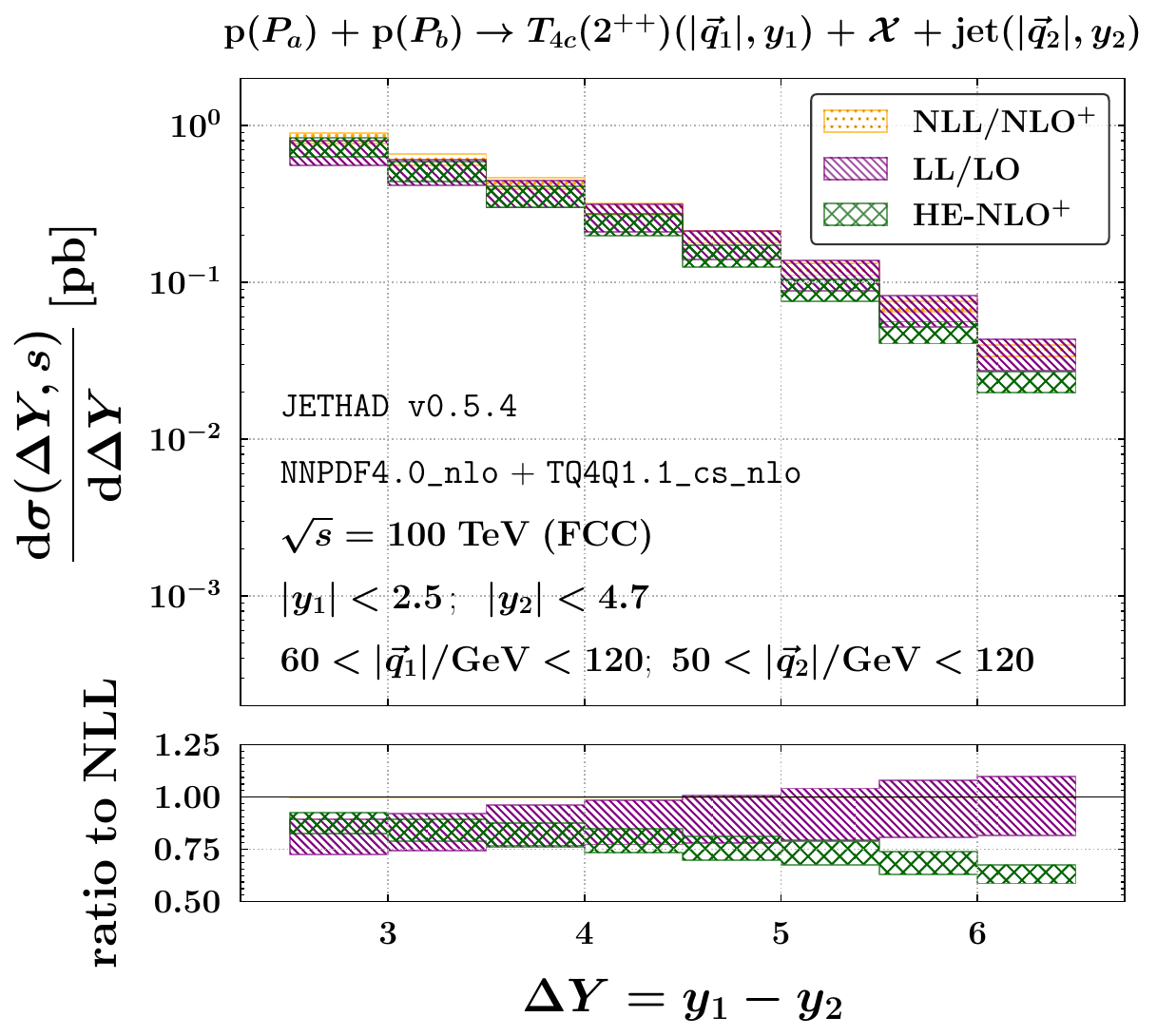}

\caption{$\DY$ differential distribution for semi-inclusive $\TQcTpp$ plus jet detections at $\sqrt{s} = 14$~TeV (HL-LHC, left) and $100$~TeV (nominal FCC, right).
{\tt NNPDF4.0} NLO proton PDFs~\cite{NNPDF:2021uiq,NNPDF:2021njg} are used in combination with {\tt TQ4Q1.1} NLO heavy-tetraquark FFs~\cite{Celiberto:2024_TQ4Q11,Celiberto:2025_TQ4Q11_AVT}.
Ancillary panels below the main plots display the ratio of $\LL$ and $\HENLOp$ predictions to $\NLLp$.
Uncertainty bands reflect the combined effect of MHOUs and numerical integration over phase space.}
\label{fig:rapidity_Tc2}
\end{figure*}

Figure~\ref{fig:rapidity_Tc2} shows the corresponding $\DY$ rates for the tensor $\TQc(2^{++})$ resonance, the most favored candidate for the $X(6900)$ state~\cite{LHCb:2020bwg}.
These plots allow for a direct comparison between two different scenarios of fully heavy-hadron production---one baryonic, the other tetraquark---in the semihard regime of QCD. 
Although both analyses share the same kinematic cuts and explore the same theoretical accuracies, the behavior of the observables is markedly different in terms of both intensity and sensitivity to the collider energy.

The ${\rm \Omega}_{3c}$ channel exhibits significantly larger cross sections, compared to the much lower rates observed for $T_{4c}$. 
Moreover, the increase in rate from 14 to 100~TeV is more pronounced for the baryon, amounting to over two orders of magnitude. 
This confirms that ${\rm \Omega}_{3c}$ production benefits more strongly from higher center-of-mass energies, and highlights the dual phenomenological character of the process at high energy.

On one hand, the LL uncertainty bands for ${\rm \Omega}_{3c}$ are more widely separated from the NLL bands than in the tetraquark case. 
They approach the full resummed prediction only gradually as $\sqrt{s}$ increases, indicating that the emergence of predictive stability under MHOUs and higher-order corrections requires higher energies for the baryon case. 
This is in contrast to $T_{4c}$, where LL and NLL bands are already closely aligned at 14 TeV, suggesting a milder sensitivity to high-energy dynamics. 
In this sense, both HL-LHC and FCC offer relevant but distinct opportunities for future studies of ${\rm \Omega}_{3c}$ plus jet emissions: the former highlights the onset of resummation effects, while the latter enables their full manifestation.

On the other hand, the $[{\rm \Omega}_{3c} + \text{jet}]$ process turns out to be much more effective than its tetraquark counterpart at discriminating the resummed signal from its fixed-order limit. 
This is a particularly important and novel result, as rapidity-interval distributions are generally considered less suited to highlighting resummation structures than transverse-momentum spectra. 
Yet in this case, the semihard production of a rare, fully heavy baryon proves to be remarkably sensitive to high-energy dynamics, setting it apart as a unique and promising probe of QCD behavior in the forward region. 
This distinctive feature underscores the relevance of ${\rm \Omega}_{3c}$ studies as a benchmark channel for future experimental and theoretical investigations.

\subsection{Transverse-momentum distributions}
\label{ssec:pT}

The second production rate matter of our phenomenological analysis is the transverse-momentum distribution
\begin{equation}
\begin{split}
\label{qT1_distribution}
 &\frac{\drv \sigma(|\vec q_1|, s)}{\drv |\vec q_1|} = 
 \int_{\DY^{\rm min}}^{\DY^{\rm max}} \drv \DY
 \int_{y_1^{\rm min}}^{y_1^{\rm max}} \drv y_1
 \int_{y_2^{\rm min}}^{y_2^{\rm max}}  \drv y_2
 \\[0.10cm]
 &\hspace{0.55cm} \times \,
 \delta (y_1 - y_2 - \DY)
 \int_{|\vec q_2|^{\rm min}}^{|\vec q_2|^{\rm max}} 
 \!\!\drv |\vec q_2|
 \, \,
 {\cal C}_{0}^{\rm [accuracy]}
 \;.
\end{split}
\end{equation}
Being differential in $|\vec q_1|$, this observable permits us to focus on the transverse-momentum spectrum of ${\rm \Omega}_{3c}$, while the jet transverse momentum is integrated in the range $40~\text{GeV} < |\vec q_2| < 120~\text{GeV}$, and $\Delta Y$ stays in a given bin of length $\DY^{\rm max} - \DY^{\rm min}$. 
The rapidity windows for both baryon and jet detections remain consistent with previous criteria.

Examining the $|\vec q_1|$ spectrum sets the groundwork for investigating potential links between the $\NLLp$ factorization approach and other theoretical frameworks. 
Allowing $|\vec q_1|$ to vary from $10$ to $100~\text{GeV}$ enables the exploration of a broad kinematic domain where additional resummation effects may become relevant. 
Recent studies on high-energy Higgs~\cite{Celiberto:2020tmb} and heavy-jet~\cite{Bolognino:2021mrc} tagging have demonstrated that our hybrid methodology is effective within the moderate transverse momentum regime, particularly when $|\vec q_1|$ and $|\vec q_2|$ are of comparable magnitude.

In contrast, higher values of $|\vec q_1|$ are expected to amplify threshold logarithms in the high-momentum domain ($|\vec q_1| \gg |\vec q_2|^{\rm max}$), while soft logarithms become significant in the low-momentum region ($|\vec q_1| \ll |\vec q_2|^{\rm min}$). 
Therefore, analyzing $|\vec q_1|$ distributions not only serves as validation for our framework, but also effectively prepares for the integration of additional resummation techniques.

\begin{figure*}[!t]
\centering

   \hspace{0.00cm}
   \includegraphics[scale=0.400,clip]{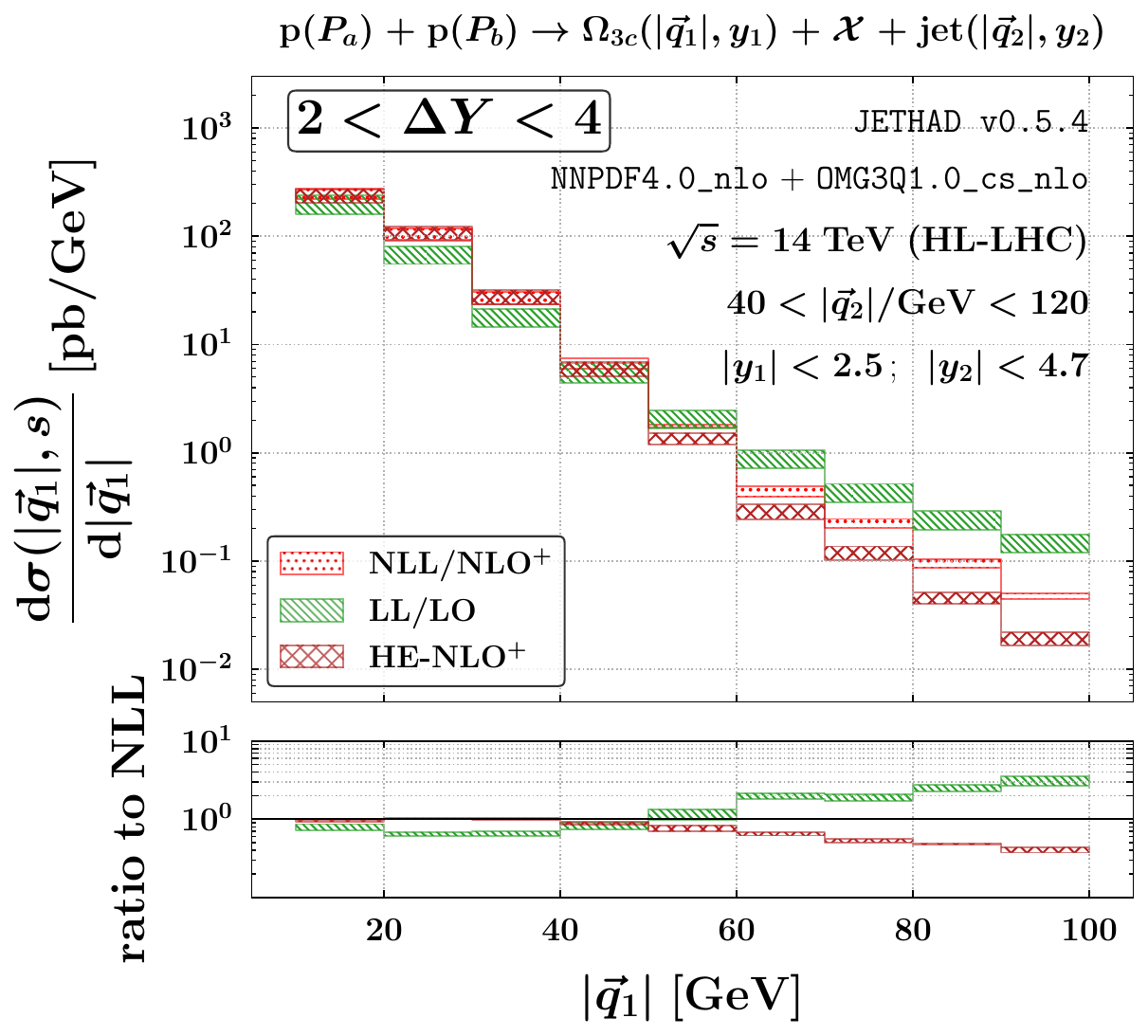}
   \hspace{0.20cm}
   \includegraphics[scale=0.400,clip]{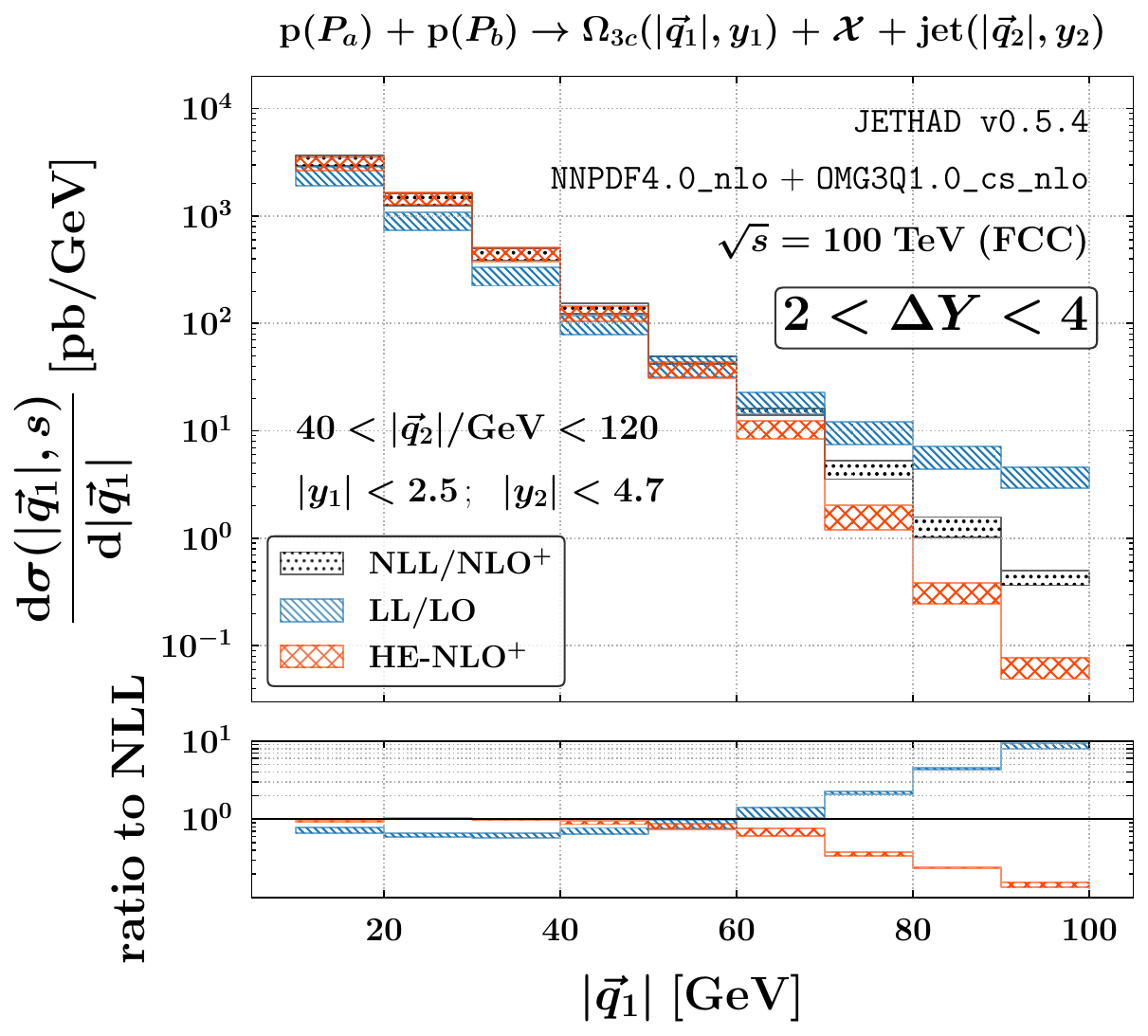}

   \vspace{0.35cm}

   \hspace{0.00cm}
   \includegraphics[scale=0.400,clip]{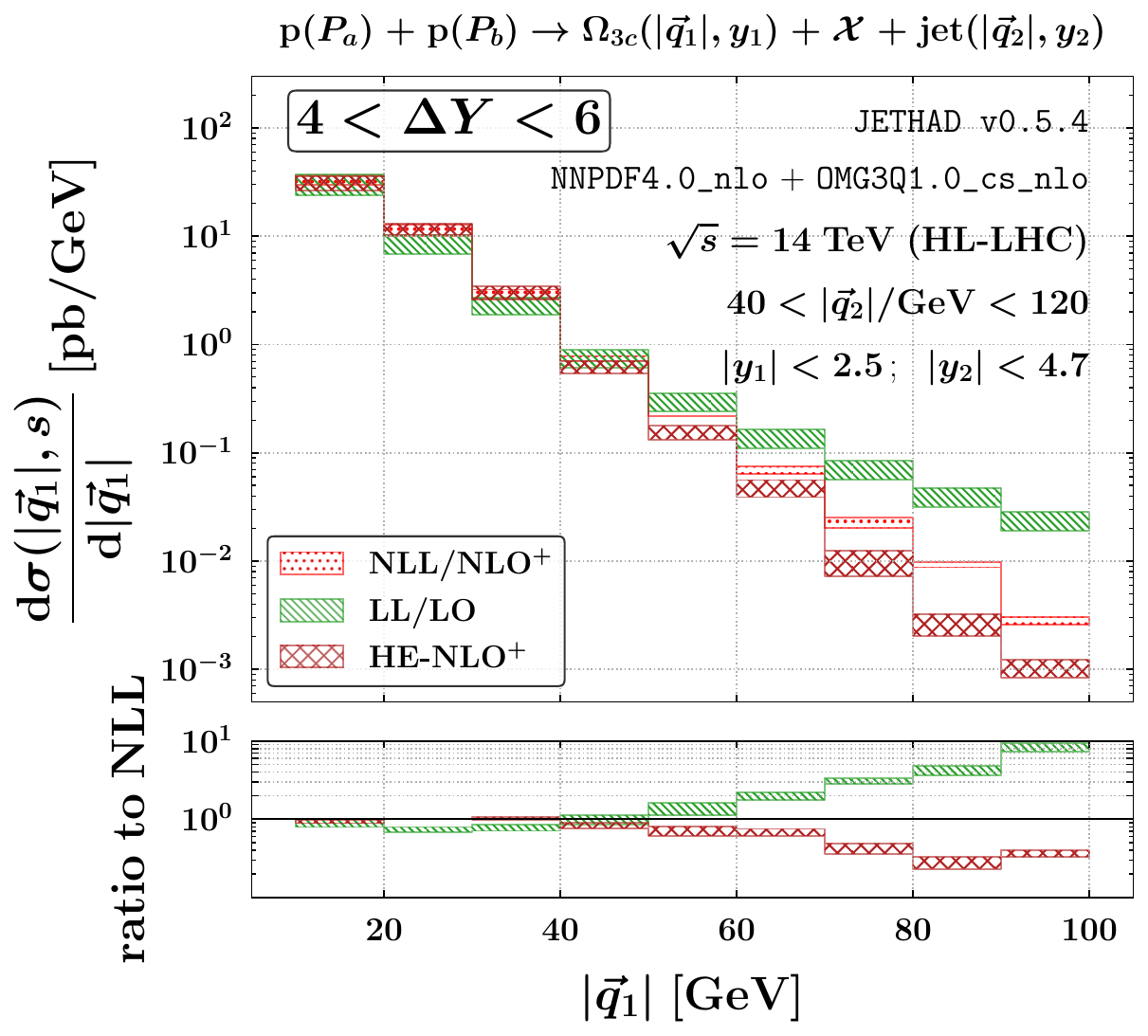}
   \hspace{0.20cm}
   \includegraphics[scale=0.400,clip]{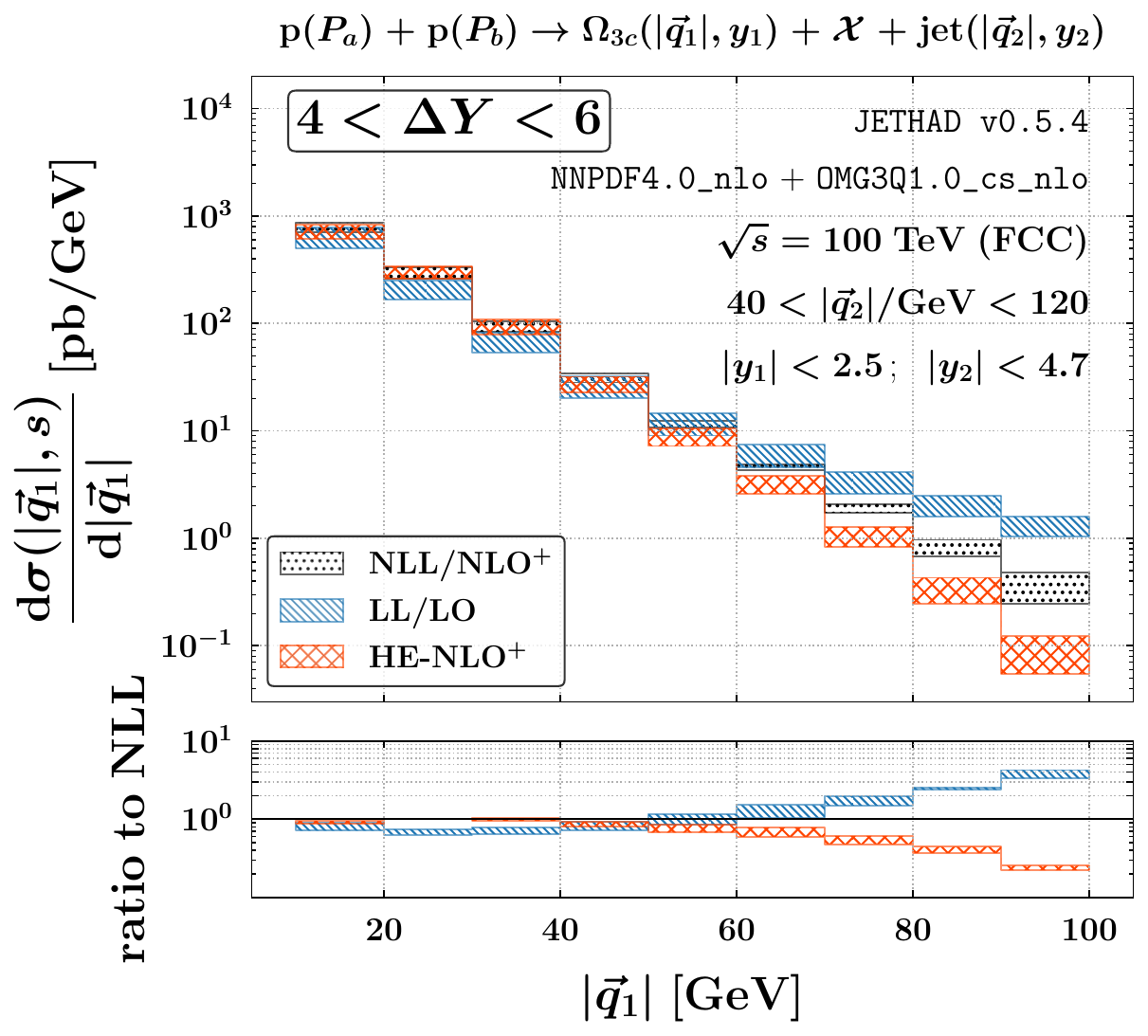}

\caption{Transverse-momentum distribution for semi-inclusive ${\rm \Omega}_{3c}$ plus jet detections at $\sqrt{s} = 14$~TeV (HL-LHC, left) and $100$~TeV (nominal FCC, right).
The rapidity interval lies within the range $2 < \DY < 4$ (upper) or $4 < \DY < 6$ (lower).
{\tt NNPDF4.0} NLO proton PDFs~\cite{NNPDF:2021uiq,NNPDF:2021njg} are used in combination with {\tt OMG3Q1.0} NLO heavy-baryon FFs~\cite{Celiberto:2025_OMG3Q10}.
Ancillary panels below the main plots show the ratio of $\LL$ and $\HENLOp$ predictions to $\NLLp$.
Uncertainty bands account for the combined effects of MHOUs and numerical phase-space integration.}
\label{fig:TM_O3c}
\end{figure*}

We present in Fig.~\ref{fig:TM_O3c} the transverse-momentum distributions of the ${\rm \Omega}_{3c}$ baryon, produced in semi-inclusive ${\rm \Omega}_{3c}$ plus jet events at $\sqrt{s} = 14$~TeV (HL-LHC, left) and $100$~TeV (nominal FCC, right). 
Results are shown for two representative rapidity intervals, $2 < \Delta Y < 4$ (top panels) and $4 < \Delta Y < 6$ (bottom panels), with each spectrum evaluated in uniform $|\vec{q}_1|$ bins of $10$~GeV. 
As expected, the overall cross sections are higher in the FCC configuration due to the extended phase-space reach, and decrease when $\Delta Y$ spans over larger values, as expected.

The $|\vec{q}_1|$ distributions exhibit a smooth and monotonic decrease as $|\vec{q}_1|$ increases, across both collider energies and $\Delta Y$ bins. 
No pronounced peak is observed; instead, the cross section steadily falls from the lowest transverse-momentum values onward. 
This behavior is consistent with expectations from collinear QCD at moderate $|\vec{q}_1|$, where the hybrid resummation framework captures the dominant logarithmic enhancements.

At $14$~TeV, the $\NLLp$ predictions show a sharper drop, reflecting stronger suppression at high $|\vec{q}_1|$ due to limited phase space and reduced parton luminosities. 
At $100$~TeV, the distributions are slightly broader and flatter, extending more visibly into the high-$|\vec{q}_1|$ region thanks to enhanced partonic fluxes. 
In all panels, the separation between the NLL signal and the two background approximations (LL and high-energy NLO) becomes more pronounced with increasing $|\vec{q}_1|$, especially at FCC energies. 
This confirms the value of transverse-momentum observables as effective probes for assessing the stability and discriminatory power of high-energy resummation in rare baryon production.

In particular, the growing discrepancy between the $\NLLp$ and $\HENLOp$ curves at larger $|\vec{q}_1|$ is not unexpected.
While BFKL resummation is well suited to configurations where $|\vec{q}_1| \simeq |\vec{q}_2|$, the regime $|\vec{q}_1| \gg |\vec{q}_2|$ falls outside its natural domain of applicability.
In this region, large logarithms of DGLAP type, together with potential threshold logarithms, become relevant, indicating the need to enhance our theoretical framework through alternative and dedicated resummation strategies tailored to these kinematic limits.

Furthermore, the behavior of the $\LL$ versus $\NLLp$ ratio reflects a complex interplay of competing NLO effects.
While jet functions typically receive negative NLO corrections~\cite{Bartels:2001ge,Ivanov:2012ms,Colferai:2015zfa}, the hadron side behaves differently: the $C_{gg}$ coefficient is corrected positively, whereas other terms yield negative shifts~\cite{Ivanov:2012iv}.
Such opposing trends can partially cancel, depending on the kinematic region, modulating the $\LL$ over $\NLLp$ behavior.
For instance, in cascade-baryon plus jet production, the ratio often exceeds unity~\cite{Celiberto:2022kxx}, unlike in doubly charmed tetraquark production~\cite{Celiberto:2023rzw}, where the gap is milder.
These variations highlight how process-dependent emission dynamics shape the relative weight of LL versus NLL contributions.

To conclude, the fair stability of our predictions under MHOU variations, as illustrated in Fig.~\ref{fig:TM_O3c}, positions the transverse-momentum spectrum of rare-baryon plus jet events as a sensitive probe of high-energy QCD mechanisms. 
This robustness, rooted in the VFNS fragmentation of ${\rm \Omega}_{3c}$, remains evident not only at current LHC scales but also at the higher energies envisioned for the FCC.

\section{Final remarks}
\label{sec:conclusions}

We explored the leading-power fragmentation of fully charmed baryons, focusing on the rare ${\rm \Omega}_{3c}$ sector at present and future hadron colliders.
To this aim, we released a new set of hadron-structure-oriented collinear FFs in a VFNS, labeled {\tt OMG3Q1.0}.
Our formulation builds on a diquark-inspired proxy model, which parallels the standard structure commonly employed in the fragmentation modeling of both singly heavy hadrons and heavy quarkonia.
In this picture, FFs are modeled as convolutions between perturbative SDCs, describing the splitting of an outgoing high-energy parton into the valence Fock state, and a purely nonperturbative component that encapsulates the hadronization process into the observed hadron.

A key feature of our method lies in the simultaneous inclusion of both charm-quark and gluon channels at the initial scale.
This dual channel structure makes ${\rm \Omega}_{3c}$ baryons an ideal testing ground for the {\HFNRevo} scheme~\cite{Celiberto:2024mex,Celiberto:2024bxu,Celiberto:2024rxa}.
Indeed, the presence of heavy partonic thresholds necessitates a dedicated DGLAP evolution strategy that treats the charm and gluon channels consistently.
By implementing a two-step evolution process---analytic below and numerical above the gluon threshold---we ensure full control over the matching between partonic sectors and the proper enforcement of threshold effects.

We evaluated the phenomenological implications of our approach using the {\psymJethad} framework.
In particular, we investigated the semi-inclusive production of ${\rm \Omega}_{3c}$ plus jet systems within a hybrid $\NLLp$ high-energy resummation scheme.
Our results, validated across HL-LHC and FCC energy scales, exhibit a progressively increasing \emph{natural stability}~\cite{Celiberto:2022grc}, directly stemming from the collinear VFNS fragmentation of rare ${\rm \Omega}$ baryons.
This stability reinforces the robustness of our formalism and underscores the predictive power of our FF determinations.

Remarkably, our phenomenological analysis has revealed the emergence of novel features.
Among these, the rapidity-interval distributions stand out for their unprecedented ability to discriminate between the NLL resummed signal and the high-energy NLO background.
Such discriminating power has not been previously observed in the context of two-object semi-inclusive hadroproduction within the semihard regime of QCD.
These findings point to a rich, nontrivial, and compelling interplay between heavy-flavor fragmentation and high-energy resummation, which opens promising avenues for targeted future studies.

Looking ahead, our goal is to enhance our framework by extending the {\HFNRevo} scheme with systematic uncertainty quantification, potentially related to MHOU effects~\cite{Kassabov:2022orn,Harland-Lang:2018bxd,Ball:2021icz,McGowan:2022nag,NNPDF:2024dpb,Pasquini:2023aaf}.
Moreover, the inclusion of additional fragmentation channels, such as those initiated by light and bottom quarks, will allow for a more complete implementation of threshold-sensitive evolution, offering further insight into the production mechanisms of rare baryons.

Heavy-flavor fragmentation remains one of the most promising intersections between hadronic structure and precision QCD.
Its relevance is amplified when studying rare or exotic systems with complex internal configurations, such as ${\rm \Omega}_{3c}$.
The unique interplay of perturbative and nonperturbative dynamics required to describe such objects deepens our understanding of hadron formation mechanisms and opens new avenues for exploring strong-interaction phenomena at high energy.

An intriguing prospect concerns the potential of rare $\rm \Omega$ baryons to serve as indirect probes of intrinsic charm content in the proton~\cite{Brodsky:1980pb,Brodsky:2015fna,Ball:2022qks,Guzzi:2022rca}. 
Because of their fully charmed valence structure, states like ${\rm \Omega}_{3c}$ provide a unique opportunity to isolate and enhance the sensitivity to initial charm contributions in high-energy collisions. 
While their production is largely influenced by gluon-initiated processes (especially at small and moderate Bjorken $x$), certain kinematic regions, such as forward rapidities and intermediate-to-large-momentum fractions, may enhance charm-initiated channels, thereby exposing potential valencelike components in the proton wave function~\cite{NNPDF:2023tyk}.

The simultaneous presence of charm and gluon initial-scale FF inputs in our {\tt OMG3Q1.0} determinations ensures that such channels are consistently treated within the {\HFNRevo} scheme. 
Despite the statistically challenging nature of ${\rm \Omega}_{3c}$ production, any anomalous enhancement in data compared to gluon-driven predictions may hint at intrinsic charm dynamics. 
In this sense, rare $\rm \Omega$ baryons could complement more conventional observables, such as $[\gamma + c]$ or $[J/\psi + c]$ systems~\cite{Flore:2020jau}, in mapping the charm content of the nucleon and in exploring its possible connection to the structure of exotic hadrons.

Our findings strengthen the case for rare baryons as precision and exploratory tools in QCD studies.
They pave the way for future experimental searches and provide a solid theoretical baseline for shedding light on the core structure hadrons, particularly in view of the far-reaching potential offered by forthcoming high-energy collider programs~\cite{Chapon:2020heu,LHCspin:2025lvj,AbdulKhalek:2021gbh,Khalek:2022bzd,Hentschinski:2022xnd,Amoroso:2022eow,Abir:2023fpo,Allaire:2023fgp,Anchordoqui:2021ghd,Feng:2022inv,AlexanderAryshev:2022pkx,LinearCollider:2025lya,LinearColliderVision:2025hlt,Arbuzov:2020cqg,Accettura:2023ked,InternationalMuonCollider:2024jyv,MuCoL:2024oxj,Black:2022cth,InternationalMuonCollider:2025sys,Accardi:2023chb}.

\section*{Acknowledgments}
\label{sec:acknowledgments}
\addcontentsline{toc}{section}{\nameref{sec:acknowledgments}}

The author thanks Alessandro Pilloni and Marco Bonvini for fruitful conversations.
The present study received support from the Atracci\'on de Talento Grant No. 2022-T1/TIC-24176 of the Comunidad Aut\'onoma de Madrid (Spain).

\section*{Data availability}
\label{sec:data_availability}
\addcontentsline{toc}{section}{\nameref{sec:data_availability}}

The {\tt LHAPDF} version of {\tt OMG3Q1.0} collinear FFs~\cite{Celiberto:2025_OMG3Q10} for fully charmed $\rm \Omega$ baryons are publicly available. They can be downloaded from the following url: \url{https://github.com/FGCeliberto/Collinear_FFs/}.



\setlength{\bibsep}{0.6em}
\bibliographystyle{apsrev}
\bibliography{references}

\end{document}